\documentclass[twocolumn,aps,superscriptaddress,showpacs,nofootinbib,floatfix]{revtex4-1}
\usepackage{epsfig,bm}
\usepackage{graphicx,amssymb}
\usepackage{amsmath}
\usepackage{epstopdf}
\usepackage[eps]{pstricks}
\usepackage[normalem]{ulem}  

\renewcommand\sout{\bgroup \color{red} \ULdepth=-.5ex \ULset}
\usepackage{slashed}
\newcommand{\comment}[1]{}
\def\Tr{{\rm{Tr}}}

\begin{document}

\title{$K_1/K^*$ enhancement as a signature of chiral symmetry restoration in heavy ion collisions}


\author{Hae-Som Sung}%
\email{ioussom@yonsei.ac.kr}
\affiliation{Department of Physics and Institute of Physics and Applied Physics, Yonsei University, Seoul 03722, Korea}

\author{Sungtae Cho}
\email{sungtae.cho@kangwon.ac.kr}
\affiliation{Division of Science Education, Kangwon National
University, Chuncheon 24341, Korea}

\author{Juhee Hong}%
\email{juhehong@yonsei.ac.kr}
\affiliation{Department of Physics and Institute of Physics and Applied Physics, Yonsei University, Seoul 03722, Korea}

\author{Su Houng Lee}%
\email{suhoung@yonsei.ac.kr}
\affiliation{Department of Physics and Institute of Physics and Applied Physics, Yonsei University, Seoul 03722, Korea}

\author{Sanghoon Lim}%
\email{shlim@pusan.ac.kr}
\affiliation{Department of Physics, Pusan National University, Pusan, Korea  }

\author{Taesoo Song}\email{t.song@gsi.de}
\affiliation{GSI Helmholtzzentrum f\"{u}r Schwerionenforschung GmbH, Planckstrasse 1, 64291 Darmstadt, Germany}


\begin{abstract}
Based on the fact that the mass difference between the chiral
partners is an order parameter of chiral phase transition and that
the chiral order parameter reduces substantially at the chemical
freeze-out point in ultra-relativistic heavy ion collisions, we
argue that the production ratio of $K_1$ over $K^*$ in such
collisions should be substantially larger than that predicted in
the statistical hadronization model. We further show that while
the enhancement effect might be contaminated by the relatively
larger decrease of $K_1$ meson than  $K^*$ meson during the
hadronic phase, the signal will be visible through a systematic
study on centrality as the kinetic freeze-out temperature is
higher and the hadronic life time shorter in peripheral collisions
than in central collisions.
\end{abstract}


\maketitle

\section{Introduction}

A recent lattice calculation finds the pseudo-critical temperature
for QCD chiral crossovers at zero baryon chemical potential to be
around $ 156.5 \pm  1.5$  MeV \cite{Bazavov:2018mes}, a value that
is consistent with previous results \cite{Borsanyi:2010bp}. This
temperature is close to the chemical freeze-out temperature
obtained from the statistical hadronization model analysis based
on the yields of hadrons produced in heavy ion collisions at both
Relativistic Heavy Ion Collider (RHIC) and Large Hadron Collider
(LHC) \cite{Andronic:2012dm, Stachel:2013zma}. Near this
temperature, the chiral order parameter associated with the
transition is substantially reduced from its vacuum value
\cite{Ding:2013lfa}. While the quark condensate is the standard
chiral order parameter, so is the difference between the vector
and axial vector correlation function. That is to say, the mass
difference between the $\rho$ and $a_1$ is due to the chiral
symmetry breaking as is well represented in the Weinberg relation
\cite{Weinberg:1967kj}. Hence, when hadrons are produced at the
chemical freeze-out point in heavy ion collisions, the chiral
partners will have masses much closer to each other than their
vacuum values. This does not mean that the masses should vanish at
the chemical freeze-out point. In fact, a recent study shows that
while chiral symmetry breaking is responsible for the mass
difference between the $\rho$ and $a_1$ meson, it only accounts
for a small fraction of the common mass \cite{Kim:2020zae}.

Statistical hadronization model analysis indicates that the
abundances of hadrons are determined at the chemical freeze-out
point \cite{Andronic:2005yp}. This suggests that the production of
chiral partners at the chemical freeze-out point will be similar.
Unfortunately, $a_1$ has a large vacuum decay width and a large
dissociation cross section so that many of the $a_1$ produced at
the chemical freeze-out will not survive during the hadronic
phase. 
Indeed, the measured yields of  resonances with large decay widths tend to be smaller than the
statistical hadronization model predictions.

On the other hand, recently it has been emphasized by one of us
that the $K_1$ and $K^*$ are also chiral partners and that both
have vacuum widths that are smaller than 100 MeV, making them ideal
objects to study the effects of partial chiral symmetry restoration in
a nuclear target experiment \cite{Song:2018plu}. In this work, we
will show that the enhancement of the ratio $K^1/K^*$ in heavy ion
collisions can be used as a signature of chiral symmetry
restoration in the early stages of heavy ion collisions. Effects
of chiral symmetry restoration in heavy ion collisions were
studied in $K^+/\pi^+$, $(\Lambda+\Sigma^0)/\pi^-$ ratios
\cite{Cassing:2015owa}. Here, we will study the $K^1/K^*$ ratio as
their properties are direct order parameters of spontaneous
chiral symmetry breaking.

\section{$K_1, K^*$ couplings in an effective Lagrangian}

From a symmetry argument alone, one expects $K_1$ to be the chiral
partner of $K^*$. In fact, arguments based solely on chiral symmetry
 predict that when chiral symmetry is restored, the vector
and axial particles become degenerate \cite{Lee:2019tvt}. For
these effects to be observable in heavy ion collisions, the vacuum
widths as well as the hadronic dissociation of both vector and
axial particles should be small so that the signal will not be
smeared out during the hadronic phase. Since the vacuum widths of
both the $K_1$ and $K^*$ meson are already smaller than 100 MeV,
we focus here on their hadronic absorptions during the hadronic
phase as well as on the centrality dependence of these effects in
a heavy ion collision. Still, in order to work out the suppression
of the initially produced $K_1$ mesons during the hadronic phase of
heavy ion collisions, we need to estimate the hadronic cross
sections with other hadrons.

For that purpose, we use the Lagrangian involving spin-1 and
spin-0 mesons following the massive Yang-Mills approach
\cite{Meissner:1987ge}:
\begin{eqnarray}\label{eq:Lagrangian1}
\mathcal{L}_{V}&=&\frac{1}{8}F^{2}_{\pi}\Tr\left(D_{\mu}U D^{\mu}U^{\dagger}\right)+\frac{1}{8}F^{2}_{\pi}\Tr\left[M\left(U + U^{\dagger}-2\right)\right]
\nonumber \\
&&-\dfrac{1}{2}\Tr\left(F_{\mu \nu L}F^{\mu \nu}_{L}+F_{\mu \nu R}F^{\mu \nu}_{R}\right)
\nonumber \\
&&+m^{2}_{0}\Tr\left(A_{\mu L}A^{\mu}_{L}+A_{\mu R}A^{\mu}_{R}\right),
\end{eqnarray}
where $ U = \exp \left(2\textit{i}P/\sqrt{2}f_{\pi}\right)$ with
the octet of pseudoscalar mesons $P = P^{a}\lambda^{a}/\sqrt{2}$
and $F_{\pi} = \sqrt{2}f_{\pi}$ = 132 MeV. The covariant
derivative and the field strength tensor are given by
\begin{eqnarray}
D_{\mu} U & = & \partial_{\mu}U -\textit{i}g\,A_{\mu\,L}U + \textit{i}g\,UA_{\mu\,R}, \nonumber \\
F_{\mu\,\nu}^{L,R} & = & \partial_{\mu}A_{\nu}^{L,R}-\partial_{\nu}A_{\mu}^{L,R}-ig\left[A_{\mu}^{L,R},A_{\nu}^{L,R}\right],
\end{eqnarray}
where $V_{\mu} = A_{\mu}^{L} + A_{\mu}^{R}$
and $ A_{\mu} =
A_{\mu}^{L} - A_{\mu}^{R} $  with the vector and axial-vector
mesons $V_{\mu} = V_{\mu}^{a}\lambda^{a}/\sqrt{2}$ and $A_{\mu}=
A_{\mu}^{a}\lambda^{a}/\sqrt{2}$.


From the first term of Eq. \eqref{eq:Lagrangian1}, we can read off
the $K^{*}K\pi$ and the $K_{1}K\rho$ interaction terms.
\begin{eqnarray}\label{eq:Lag3}
\mathcal{L}^{(3)} &=&\,\frac{\textit{i}g}{2}\Tr\left(V_{\mu}P\overleftrightarrow{\partial}^{\mu}P\right)
\nonumber \\
&&+\frac{\textit{i}g^{2}F_{\pi}}{4}\Tr\left(V_{\mu}P A^{\mu}-V_{\mu} A^{\mu}P\right)
\nonumber \\
&&+\frac{\textit{i}g}{2}\Tr\left[\left(\partial_{\mu} V_{\nu}-\partial_{\nu} V_{\mu}\right)V^{\mu}V^{\nu}\right]
\nonumber \\
&&+\textit{i}g\Tr\left[\left(\partial_{\mu} V_{\nu}-\partial_{\nu} V_{\mu}\right)A^{\mu}A^{\nu}-\partial_{\nu}A_{\mu}A^{\mu}V^{\nu}\right].
\end{eqnarray}
Specifically,
\begin{eqnarray}
\mathcal{L}_{VPP}
&=&  \frac{\textit{i}g}{2}\bar{K}^{*\,\mu}\left[\left(\vec{\tau}\cdot\partial_{\mu}\vec{\pi}\right)K-\left(\vec{\tau}\cdot\vec{\pi}\right)\partial_{\mu}K\right]
\nonumber \\
&& +\frac{\textit{i}g}{2}\bar{K}\left[\left(\vec{\tau}\cdot\vec{\rho}^{\mu}\right)\right]\partial_{\mu}K + H.C.,\label{eq:LagVPP}
 \\
\mathcal{L}_{AVP} 
&=& \textit{i}\frac{g^{2}F_{\pi}}{4}\bar{K_{1}}^{\mu}\left[\left(\vec{\tau}\cdot\vec{\rho_{\mu}} \right)K - \left(\vec{\tau}\cdot\vec{\pi}\right)K^{*}_{\mu}\right] 
+ H.C.,
\nonumber \\ \label{eq:LagAVP}
\end{eqnarray}
with $K \equiv (K^{+},K^{0})^{T}$ ,
$K^{*}\equiv(K^{+},K^{*0})^{T}$ and $K_{1}\equiv(K^{+}_{1},
K^{0}_{1})^{T}$ being the strange pseudoscalar, vector and
axial-vector meson isospin doublets, respectively. $\vec{\pi}$ and
$\vec{\rho}$, with Pauli matrices $\vec{\tau}$, are the pion and
$\rho$ meson isospin triplets. 

As it stands, the interaction terms for $K_{1}K^{*}\pi$ and
$K_{1}K\rho$ have a common coupling as given from Eq.
\eqref{eq:Lag3}. This leads to a larger calculated decay width of
$K_{1}K^{*}\pi$ compared with that of $K_{1}K\rho$, because the
phase space for the former two-body decay is larger than that for 
the latter. On the other hand, experimentally one finds
$\Gamma_{K_{1}\rightarrow K^{*}\pi} = 25$ MeV and
$\Gamma_{K_{1}\rightarrow K \rho} = 38$ MeV. The resolution of
this puzzles comes from mixing effects. In the vacuum, the
low-lying modes that couple to the vector current are $K^*(892)$
and $K^*(1410)$ while those for the axial vector current are
$K_1(1270)$ and $K_1(1400)$. There is a subtlety in the nature of
the two $K_1$ states: they are assumed to be a mixture of the
$^3P_1$ and $^1P_1$ quark-antiquark pair in the quark model
\cite{Suzuki:1993yc}. We therefore introduce two mixing angles
$\theta_A$ and $\theta_V$ so that the chiral Lagrangian in Eq. \eqref{eq:LagVPP} and \eqref{eq:LagAVP} are 
constructed for both $(K^*_A,K_{1A})$ and $(K^*_B,K_{1B})$ that
are related to the physical mesons as follows:
\begin{eqnarray}
K_{1 A} &=& \cos\theta_{A}K_{1}(1270) + \sin\theta_{A}K_{1}(1400),
\nonumber \\
K_{1 B} &=& -\sin\theta_{A}K_{1}(1270) + \cos\theta_{A}K_{1}(1400),
\nonumber \\
\nonumber \\
K^{*}_{A} &=& \cos\theta_{V}K^{*}(892) +\sin\theta_{V}K^{*}(1410),
\nonumber \\
K^{*}_{B} &=& -\sin\theta_{V}K^{*}(892) +\cos\theta_{V}K^{*}(1410).
\end{eqnarray}

Now, we re-express the sum of the two chiral Lagrangians in terms of
$K_{1}$ and $K^*$ mesons and the two independent couplings $g_A, g_B$
so that the SU(2) chiral symmetry is still preserved.
\begin{eqnarray} 
\mathcal{L}^{'}_{VPP}&=&\textit{i}g_{K^{*}K\pi}\,\bar{K}^{*\,\mu}\left[\left(\sqrt{2}\vec{\tau}\cdot\partial_{\mu}\vec{\pi}\right)K -\left(\sqrt{2}\vec{\tau}\cdot\vec{\pi}\right)\partial_{\mu}K\right]
\nonumber \\
&&+H.C.,  \label{six-decay1}
\\
\mathcal{L}^{'}_{AVP}  &=& \textit{i}g_{K_{1}K\rho}\bar{K_{1}}^{\mu}\left(\sqrt{2}\vec{\tau}\cdot\vec{\rho}_{\mu}\right)K 
\nonumber \\
&&-\textit{i}g_{K_{1}K^{*}_{\mu}\pi}\bar{K_{1}}^{\mu}\left(\sqrt{2}\vec{\tau}\cdot\vec{\pi}\right)K^{*}_{\mu} + H.C., \label{six-decay2}
\end{eqnarray}
where
\begin{eqnarray}
g_{K^{*}K\pi} &=& \frac{1}{4\sqrt{2}}\left(g_{A}\cos\theta_{V} - g_{B}\sin\theta_{V}\right),
\nonumber \\
g_{K^{*}(1410)K\pi} &=& \frac{1}{4\sqrt{2}}\left(g_{A}\sin\theta_{V} + g_{B}\cos\theta_{V}\right),
\nonumber \\
\nonumber \\
g_{K_{1}K\rho} &=& \frac {F_{\pi}}{8\sqrt{2}}\left(g_{A}^{2}\cos\theta_{A} - g_{B}^{2}\sin\theta_{A}\right),
\nonumber \\
g_{K_{1}(1400)K\rho} &=& \frac {F_{\pi}}{8\sqrt{2}}\left(g_{A}^{2}\sin\theta_{A} + g_{B}^{2}\cos\theta_{A}\right),
\nonumber \\
\nonumber \\
g_{K_{1}K^{*}\pi} &=& \frac {F_{\pi}}{8\sqrt{2}}\left(g_{A}^{2}\cos\theta_{A}\cos\theta_{V} + g_{B}^{2}\sin\theta_{A}\sin\theta_{V}\right),
\nonumber \\
g_{K_{1}(1400)K^{*}\pi} &=& \frac {F_{\pi}}{8\sqrt{2}}\left(g_{A}^{2}\sin\theta_{A}\cos\theta_{V} - g_{B}^{2}\cos\theta_{A}\sin\theta_{V}\right).
\nonumber \\
\end{eqnarray}

Using the above interaction Lagrangians, we determine the mixing
angle $\theta_{A}$($\theta_{B}$) and coupling constant
$g_{A}$($g_{B}$) to best fit the experimental data for the six
decay processes given in Table
\ref{table:Coupling}.  The table also summarizes the best fit couplings and the
experimental data as well as the calculated decay widths of the
six processes.
\begin{center}
\begin{table*}[htbp]
 \begin{tabular}{ccccccc}
 \hline  \hline
ABC &
$K^{*}(892) K\pi$ & $K^{*}(1410)K\pi$ & $K_{1}(1270)K\rho$ & $K_{1}(1400)K\rho$  &  $K_{1}(1270)K^{*}(892)\pi$ & $K_{1}(1400)K^{*}(892)\pi$
  \\ [0.5ex]
   \hline
$g_{ABC}$  & 3.17 & 0.9 & 2.9 (GeV) & $-1.02$ (GeV) &0.649 (GeV) &$-1.95$ (GeV)
 \\ \hline
$\Gamma_{A \rightarrow BC}$(experiment)  &50.0 (MeV)& 15.6 (MeV) & 37.8 (MeV) & 5.2 (MeV) & 14.4 (MeV) &163 (MeV)
 \\ \hline
$\Gamma_{A \rightarrow BC}$(calculated) &48.4 (MeV)& 14.8 (MeV) & 39.7 (MeV) & 14.7 (MeV) & 9.42 (MeV) &98.6 (MeV)
 \\ \hline
\end{tabular}
\caption{The coupling constants are calculated with  $g_{A}$=15.5, $g_{B}$= -10.4, $\cos\theta_{V}$=0.6469, $\sin\theta_{V}$=0.7626, $\cos\theta_{A}$=0.724 and $\sin\theta_{A}$= -0.6898. With these parameters we also  find four point couplings $g_{K_{1}(1270)K\pi\pi}=$7.67(GeV$^{-1}$) and $g_{K_{1}(1400)K\pi\pi}=-34.5$(GeV$^{-1}$). }
\label{table:Coupling}
\end{table*}
\end{center}

\section{The cross section of the $K_{1}$ meson}

We assume that $K_{1}$ mesons are in thermal equilibrium when they
are produced at the chemical freeze-out. Then $K_{1}$ mesons
interact with other particles until the kinetic freeze-out point. We
consider the hadronic effect on the $K_{1}$ meson due to
interactions with light mesons such as pions and rho mesons;
$K_{1} + \pi \rightarrow K + \pi$, $K_{1} + \pi \rightarrow K^{*}
+ \rho$,  $K_{1} + \rho \rightarrow  K^{*} + \pi$, and $K_{1} +
\rho \rightarrow K + \rho$. In addition to the interaction terms given in Eq. \eqref{six-decay1} and \eqref{six-decay2}, the Lagrangian describing the
interaction between the $K_{1}$ meson and pions and rho mesons is
given as follows:

\begin{eqnarray}\label{eq:APPP} 
\nonumber
\mathcal{L}^{'}_{APPP} &=& 2 g_{K_{1}K\pi\pi}\bar{K_{1}}^{\mu}\left[\left(\vec{\tau}\cdot\partial_{\mu}\vec{\pi}\right)\left(\vec{\tau}\cdot\vec{\pi}\right)K +\left(\vec{\tau}\cdot\vec{\pi}\right)^{2}\partial_{\mu}K\right],
\nonumber \\
\mathcal{L}^{'}_{VPP} &=& \textit{i}g_{\rho\pi\pi}[\rho^{0\,\mu}(\partial_{\mu}\pi^{+}\,\pi^{-}-\pi^{+}\partial_{\mu}\pi^{-})
\nonumber \\
&+&\rho^{+\,\mu}(\partial_{\mu}  \pi^{-}\,\pi^{0}-\pi^{-}\partial_{\mu}\pi^{0})
\nonumber \\
&+& \rho^{-\,\mu}(\partial_{\mu}\pi^{0}\,\pi^{+}-\pi^{0}\partial_{\mu}\pi^{+})] +H.C.
\nonumber \\
&+&\textit{i}g_{KK\rho}\left[\bar{K}\left(\sqrt{2}\vec{\tau}\cdot\vec{\rho}^{\mu}\right)\partial_{\mu}K-\partial_{\mu}\bar{K}\left(\sqrt{2}\vec{\tau}\cdot\vec{\rho}^{\mu}\right)K\right],
\nonumber \\
\mathcal{L}^{'}_{AVP} &=& \textit{i}g_{a_{1}\rho\pi}\lbrace\rho^{+\,\mu}(\pi^{-}a_{1\,\mu}^{0}-\pi^{0}a_{1\,\mu}^{-})
\nonumber \\
&+&\rho^{-\,\mu}(\pi^{0}a_{1\,\mu}^{+}-\pi^{+}a_{1\,\mu}^{0})+\rho^{0\,\mu}(\pi^{+}a_{1\,\mu}^{-}-\pi^{-}a_{1\,\mu}^{+})\rbrace
\nonumber \\
\mathcal{L}^{'}_{VVV} &=& \sqrt{2}\textit{i}g_{K^{*}K^{*}\rho}[\bar{K^{*}}^{\nu}\vec{\tau} K^{* \,\mu}(\partial_{\mu}\vec{\rho}_{\nu}-\partial_{\nu}\vec{\rho}_{\mu})
\nonumber \\
&+&(\partial_{\mu}\bar{K^{*}}_{\nu}-\partial_{\nu}\bar{K^{*}}_{\mu})\vec{\tau}\cdot\vec{\rho}^{\mu}K^{*\,\nu}
\nonumber \\
&+&\bar{K^{*}}^{\mu}\vec{\tau}\cdot\vec{\rho}^{\nu}(\partial_{\mu}K^{*}_{\nu}-\partial_{\nu}K^{*}_{\mu})]
\nonumber \\
&+&\frac{\textit{i}g}{2}[(\partial_{\mu}(\vec{\tau}\cdot\vec{\rho}_{\nu})-\partial_{\nu}(\vec{\tau}\cdot\vec{\rho}_{\mu}))(\vec{\tau}\cdot\vec{\rho}^{\mu})(\vec{\tau}\cdot\vec{\rho}^{\nu})],
\nonumber \\
\mathcal{L}^{'}_{AAV} &=& \textit{i}\sqrt{2}g_{K_{1}a_{1}K^{*}}[\partial_{\nu}\bar{K}_{1\,\mu}\lbrace(\vec{a}^{\nu}_{1}\cdot\vec{\tau})K^{*\,\mu}
\nonumber \\
&-&(\vec{a}^{\mu}_{1}\cdot\vec{\tau})K^{*\,\nu}\rbrace
-\bar{K}_{1}^{\mu}\partial_{\mu}(\vec{a}_{1\,\nu}\cdot\vec{\tau})K^{*\,\nu}]
\nonumber \\
&+&\textit{i}\sqrt{2}g_{K_{1}K_{1}\rho}[\lbrace\bar{K}_{1}^{\mu}(\vec{\rho}^{\nu}\cdot\vec{\tau})-\bar{K}_{1}^{\nu}(\vec{\rho}^{\mu}\cdot\vec{\tau})\rbrace\partial_{\nu}K_{1\,\mu}
\nonumber \\
&-&\textit{i}g\partial_{\mu}\bar{K}_{1}^{\nu}(\vec{\rho}_{\nu}\cdot\vec{\tau})K_{1}^{\mu}].
\end{eqnarray}
The coupling constant $g=8.49$ is determined by fitting the
experimental decay width of $\rho \rightarrow \pi\pi$, from which we find
$g_{\rho\pi\pi} = g_{\rho\rho\rho}=g/\sqrt{2}=6.006$ and
$g_{KK\rho} = g/2\sqrt{2} = 3.003$.  In the strange vector and
axial-vector mesons, we apply the mixing angle and coupling
constants $g_{A}$ and $g_{B}$, which were determined to give the coupling
constants as given in Table \ref{table:Coupling} and additionally $g_{K^{*}K^{*}\rho} = 0.078$,
$g_{K_{1}K_{1}\rho}=1.12$, $g_{K_{1}a_{1}K^{*}}=4.5$ and
$g_{a_{1}\rho\pi} =3.36$ (GeV).

\begin{figure*}
\begin{minipage}{0.23\textwidth}
\begin{center}
\includegraphics[height=2.2cm,width =2.6cm]{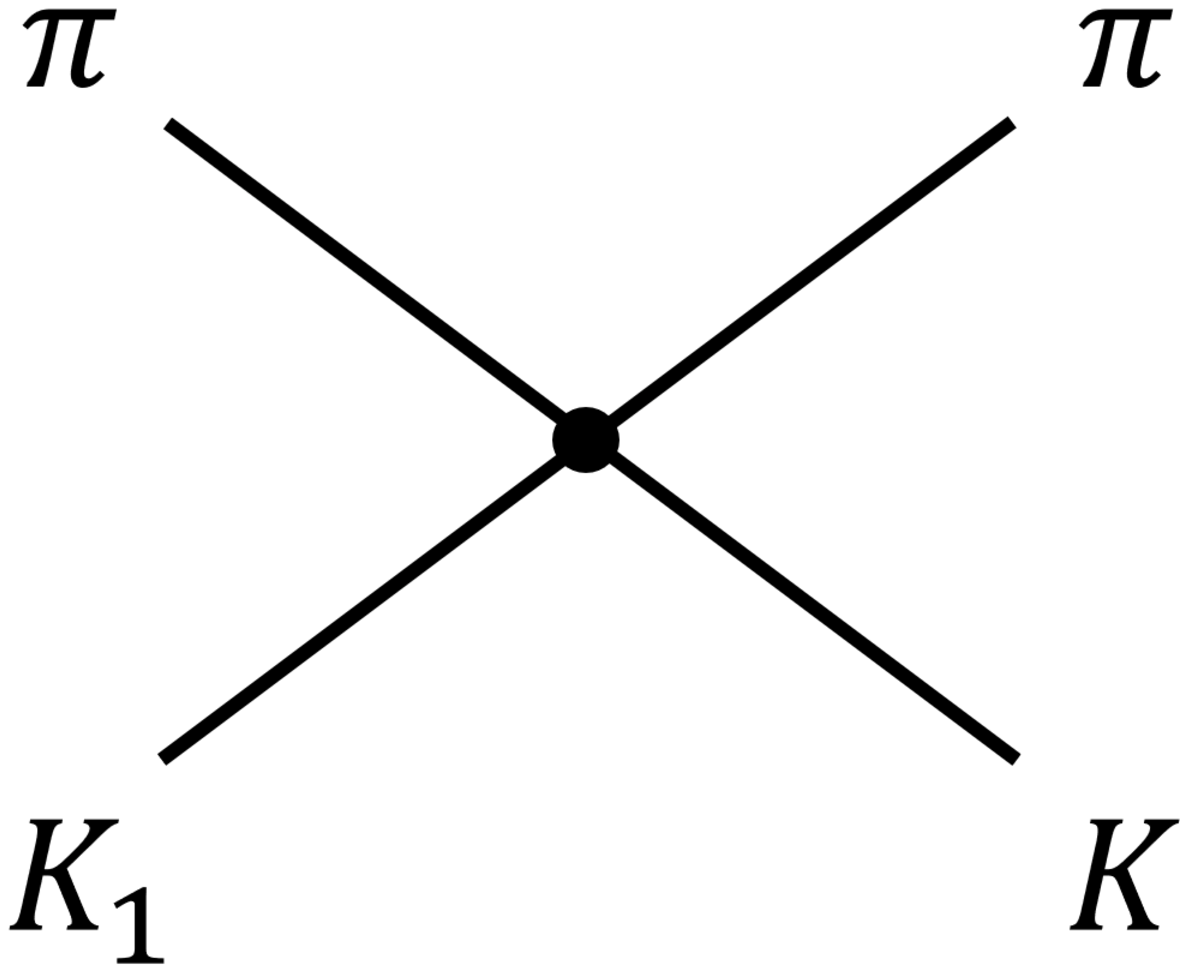}
\end{center}
\end{minipage}
\begin{minipage}{0.24\textwidth}
    \includegraphics[height=2.2cm]{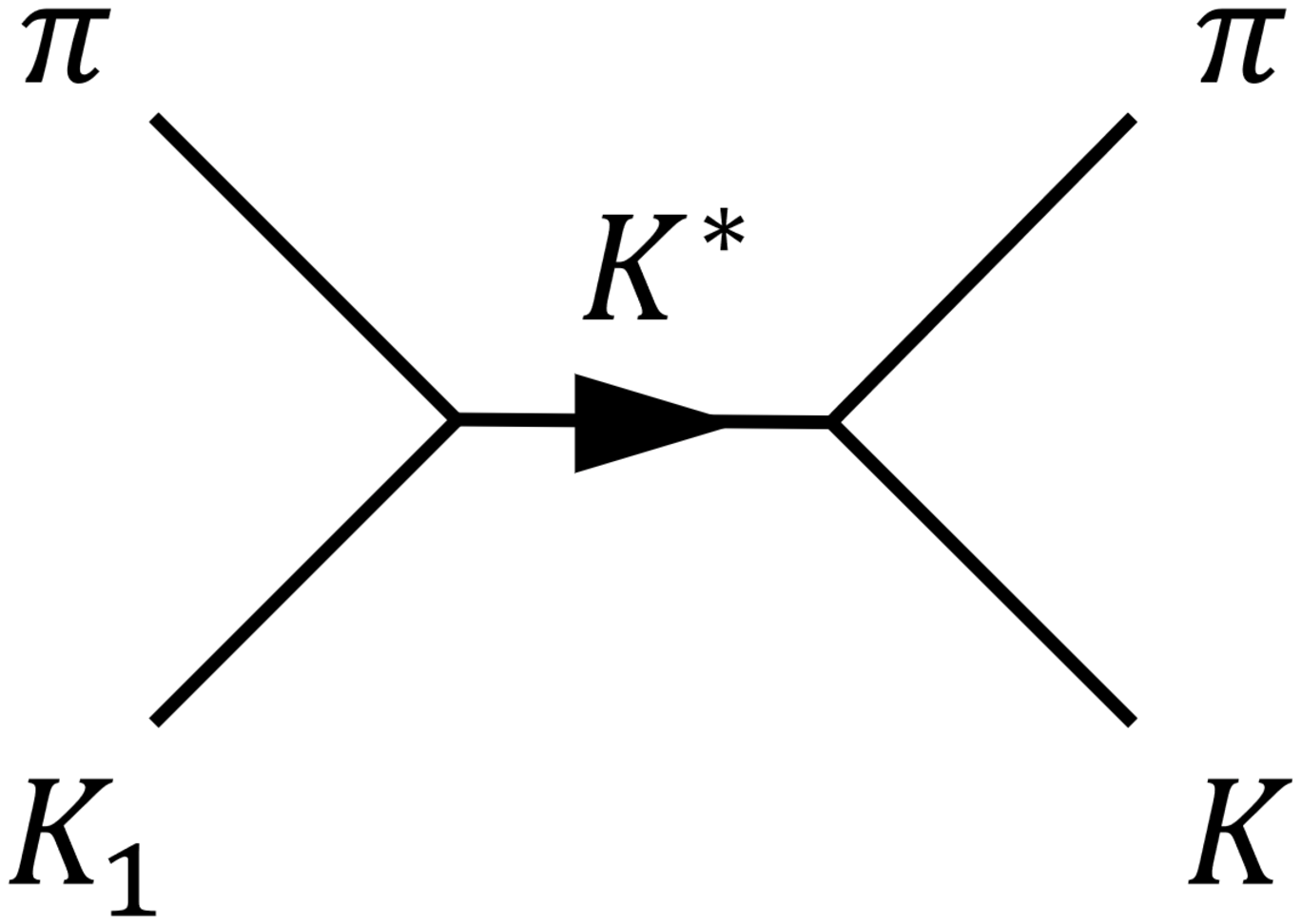}
\end{minipage}
\begin{minipage}{0.24\textwidth}
    \includegraphics[height=2.2cm]{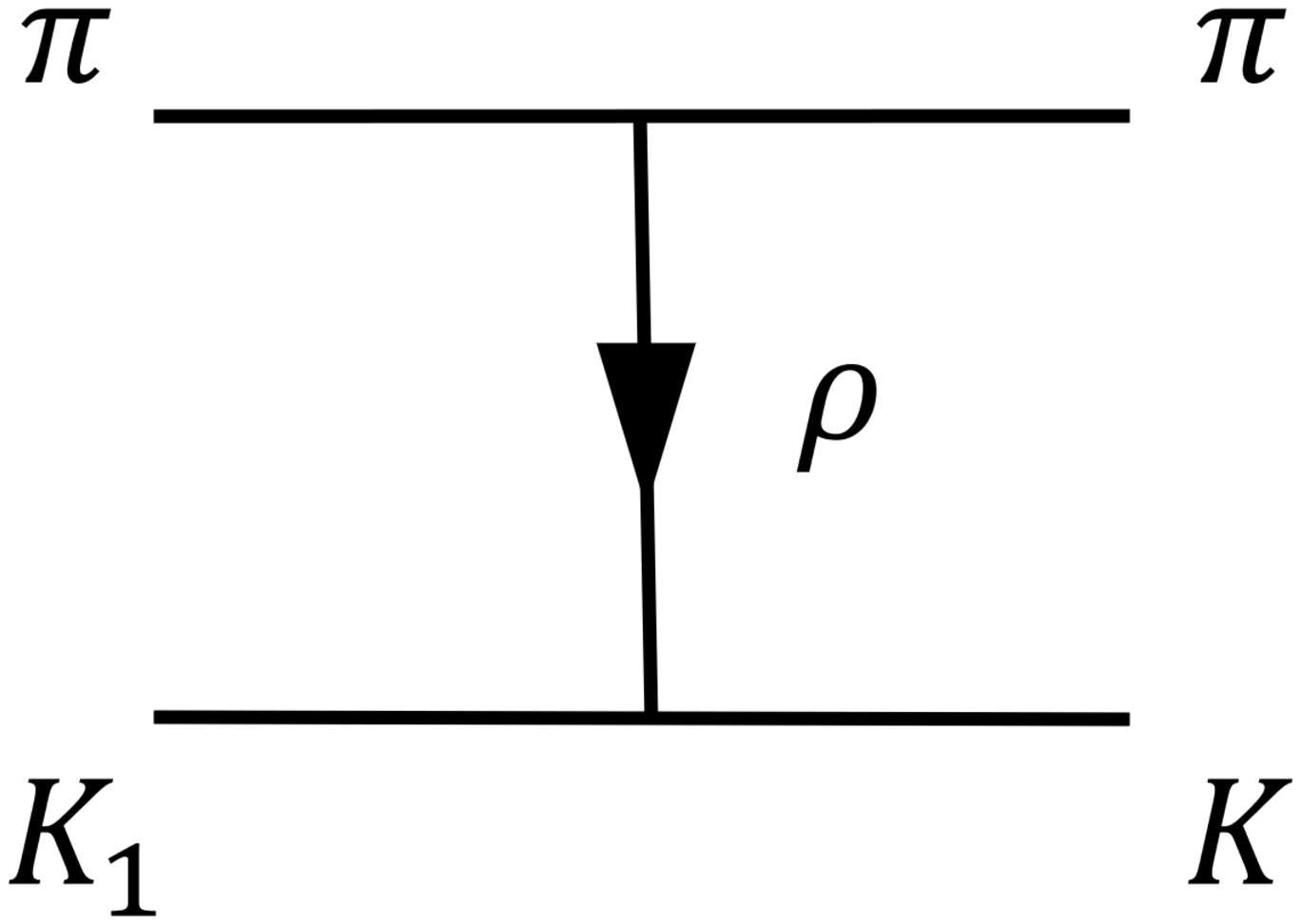}
\end{minipage}
\begin{minipage}{0.25\textwidth}
    \includegraphics[height=2.2cm]{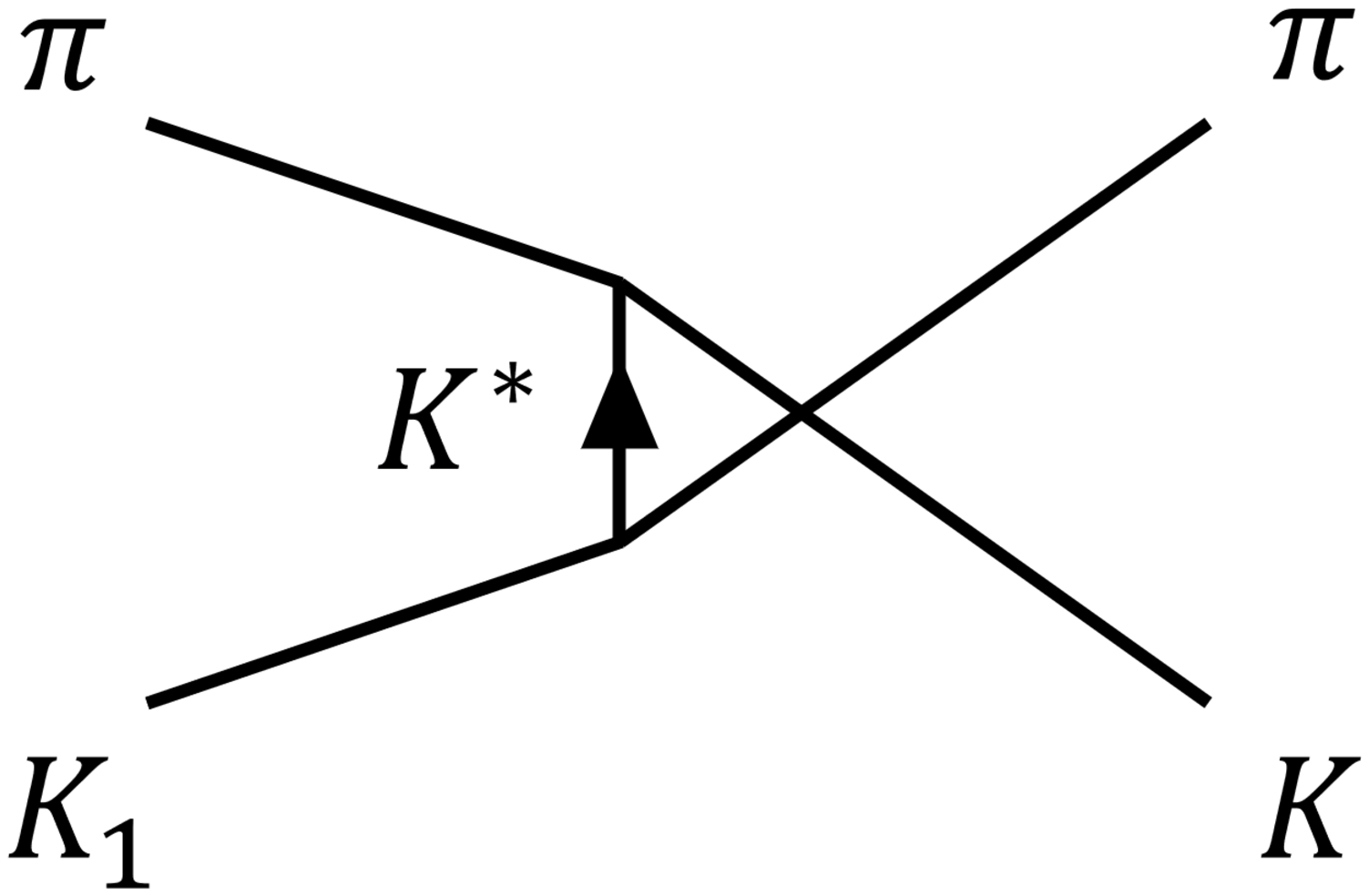}
\end{minipage}
\vspace{0.2cm}
\begin{minipage}{0.23\textwidth}
(1-a)
\end{minipage}
\begin{minipage}{0.24\textwidth}
(1-b)
\end{minipage}
\begin{minipage}{0.24\textwidth}
(1-c)
\end{minipage}
\begin{minipage}{0.25\textwidth}
(1-d)
\end{minipage}
\vspace{0.2cm}
\vspace{0.05cm}
\begin{minipage}{0.18\textwidth}
\includegraphics[height=1.9cm]{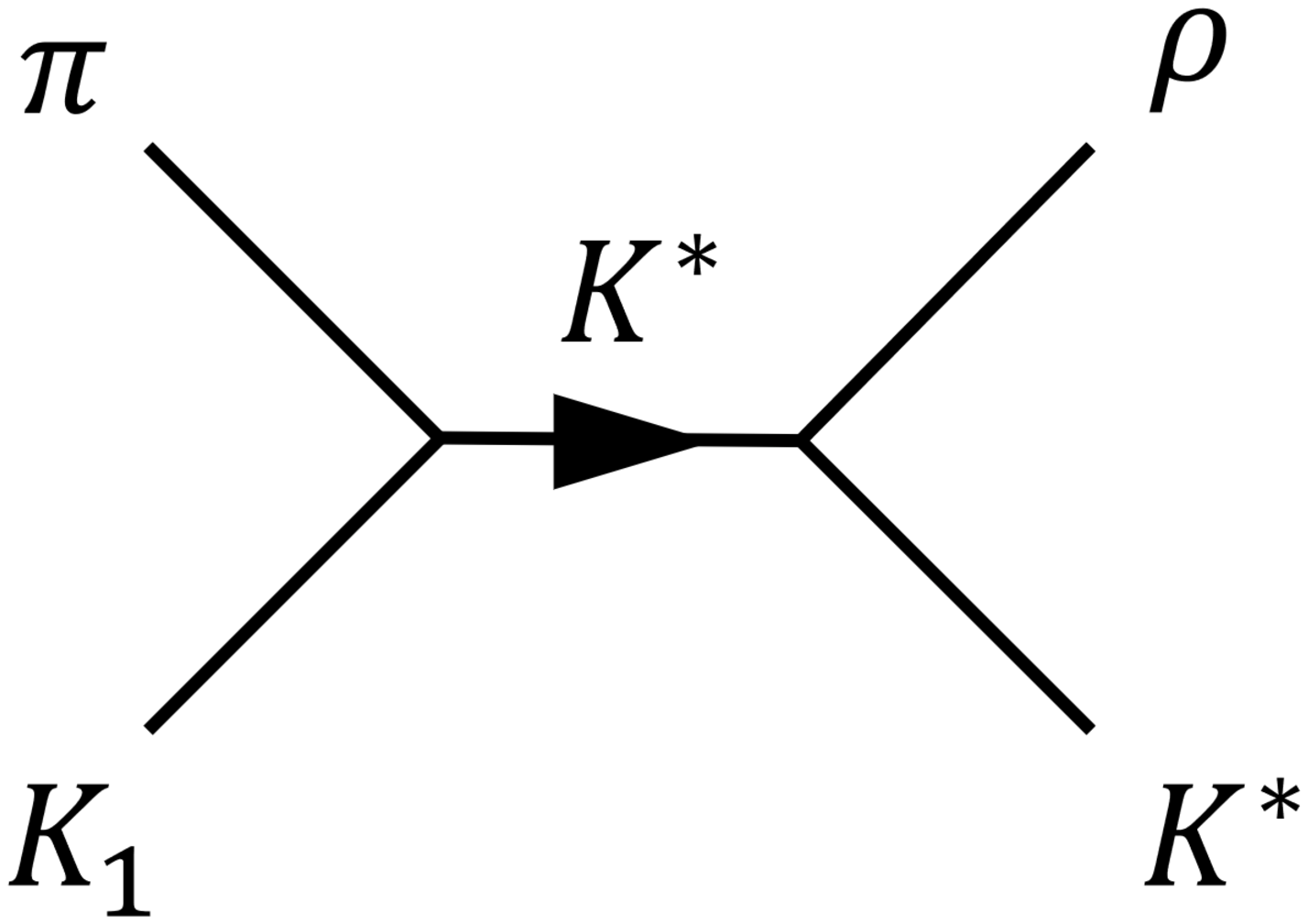}
\end{minipage}
\begin{minipage}{0.18\textwidth}
    \includegraphics[height=1.9cm]{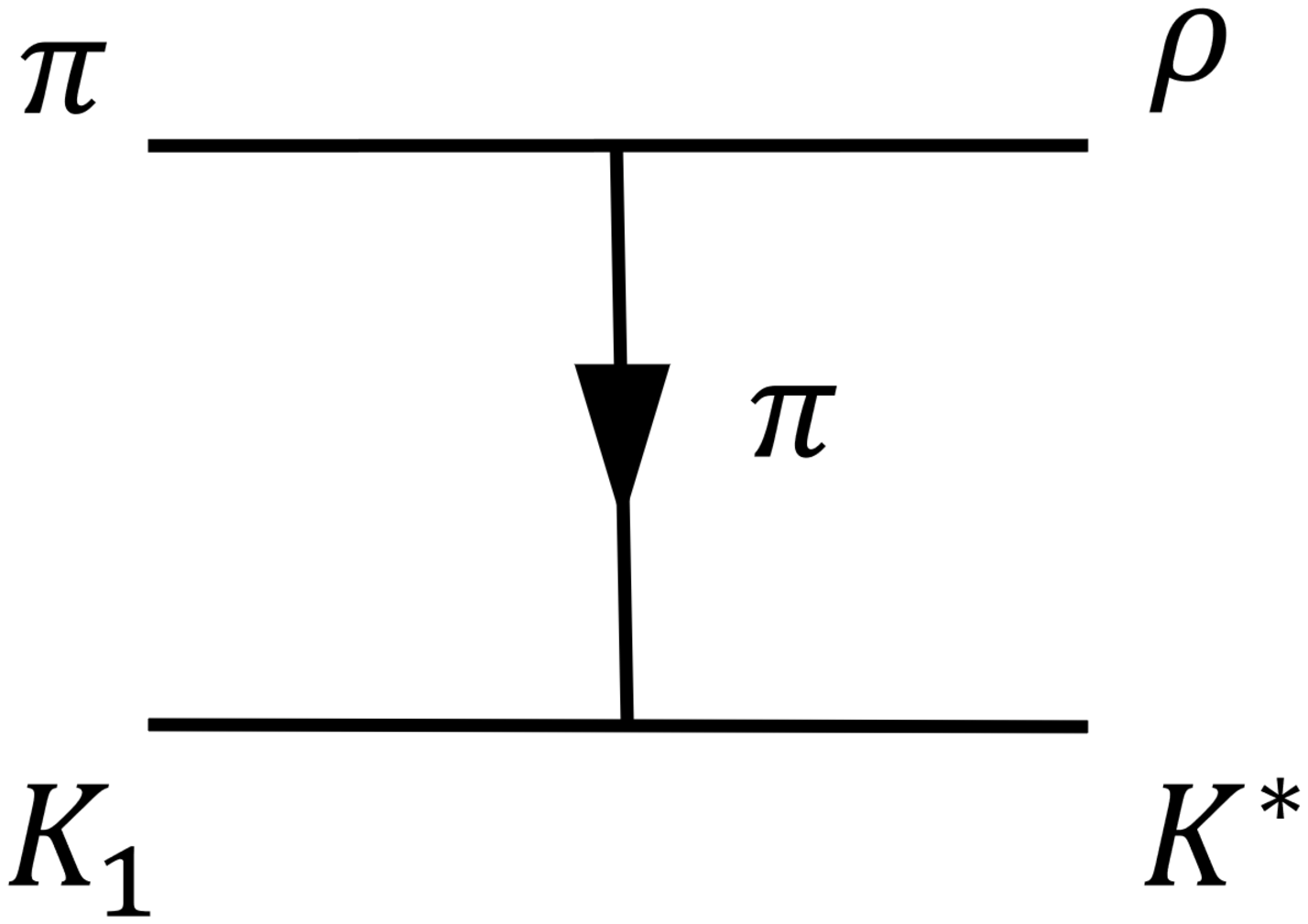}
\end{minipage}
\begin{minipage}{0.19\textwidth}
\includegraphics[height=1.9cm]{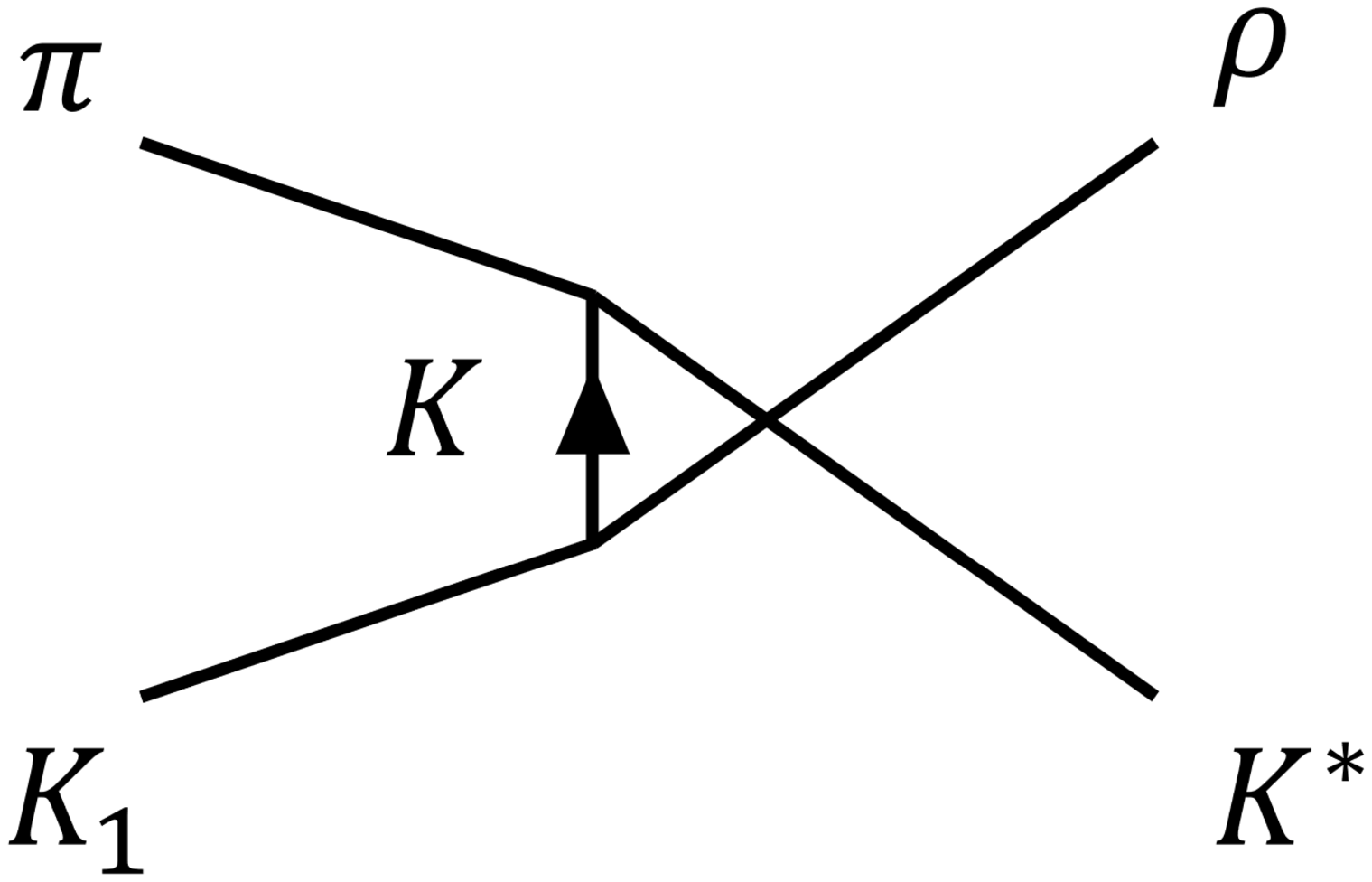}
\end{minipage}
\begin{minipage}{0.18\textwidth}
    \includegraphics[height=1.9cm]{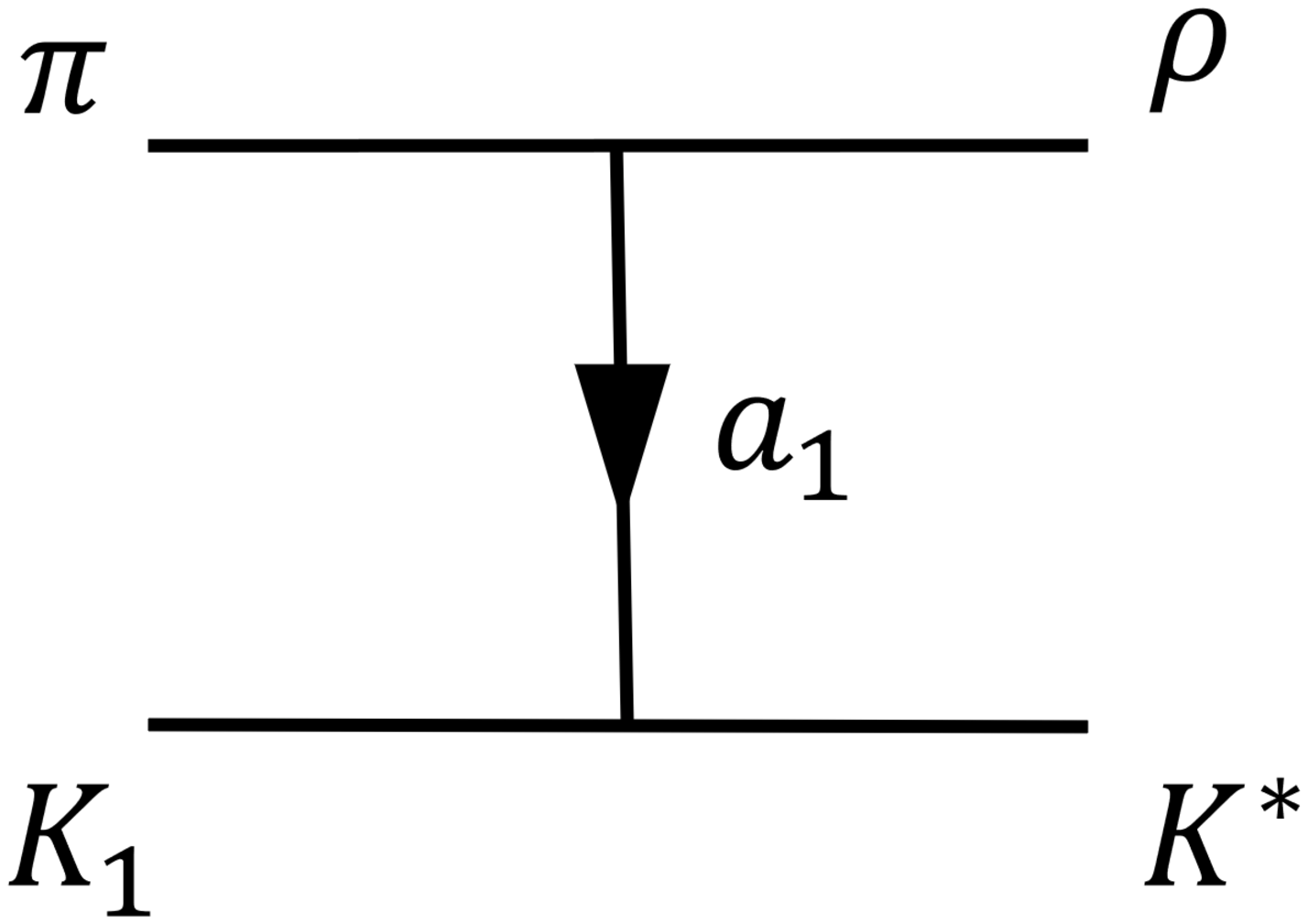}
\end{minipage}
\begin{minipage}{0.19\textwidth}
\includegraphics[height=1.9cm]{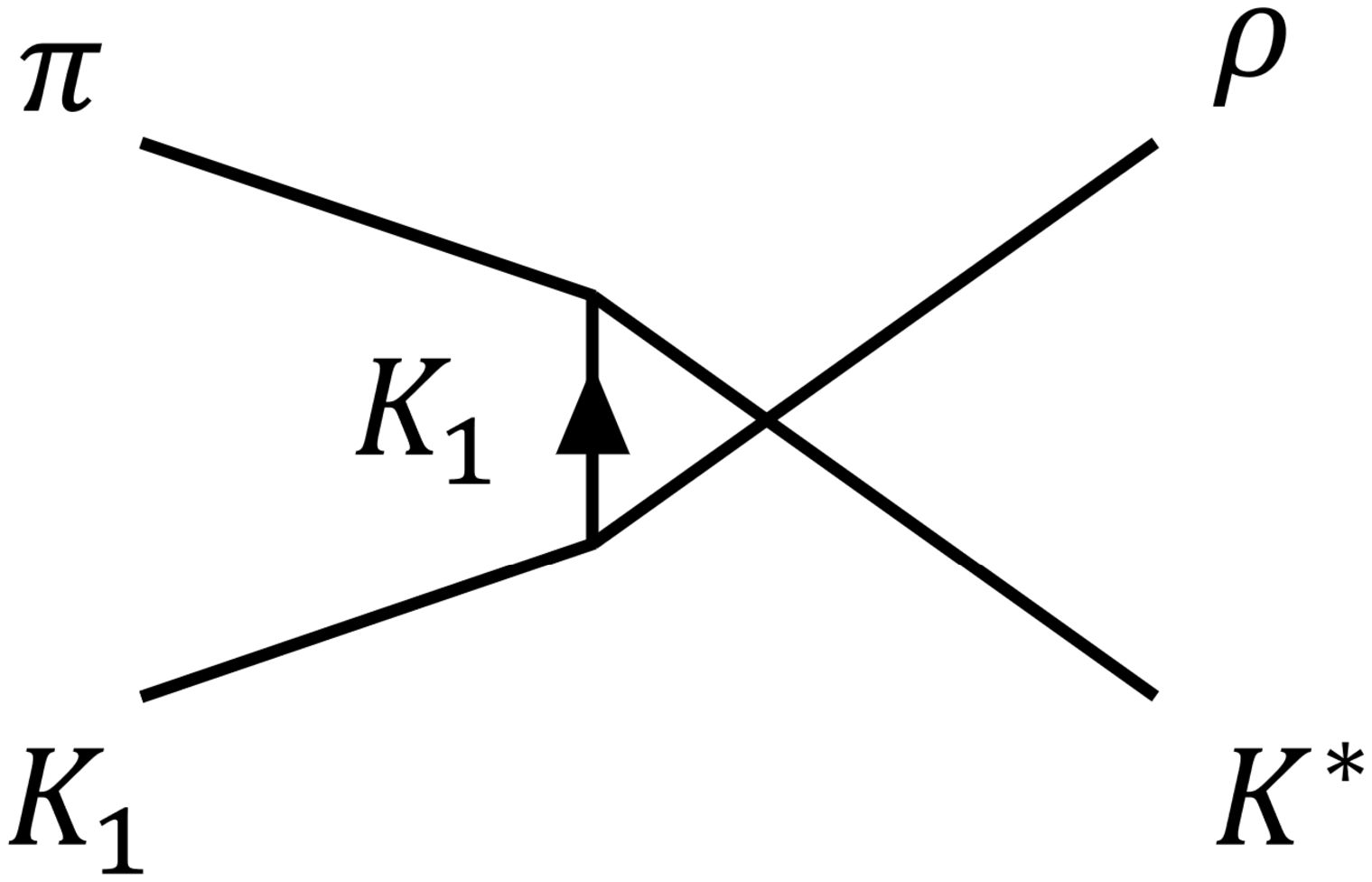}
\end{minipage}
\vspace{0.2cm}
\begin{minipage}{0.18\textwidth}
(2-a)
\end{minipage}
\begin{minipage}{0.18\textwidth}
(2-b)
\end{minipage}
\begin{minipage}{0.19\textwidth}
(2-c)
\end{minipage}
\begin{minipage}{0.18\textwidth}
(2-d)
\end{minipage}
\begin{minipage}{0.19\textwidth}
(2-e)
\end{minipage}
\vspace{0.2cm}
\vspace{0.05cm}
\begin{minipage}[t]{0.24\textwidth}
    \includegraphics[height=2.3cm]{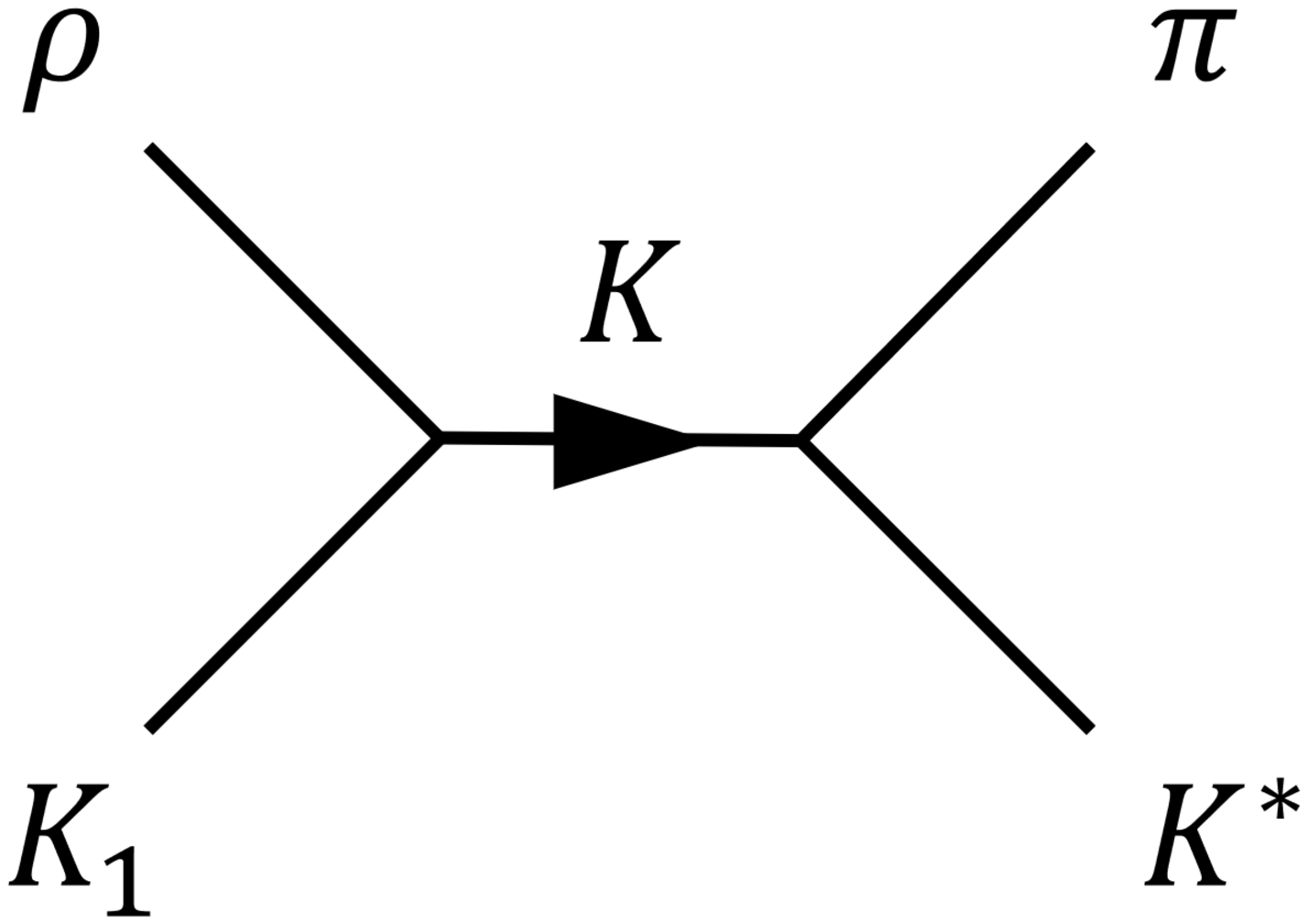}
\end{minipage}
\begin{minipage}[t]{0.24\textwidth}
    \includegraphics[height=2.3cm]{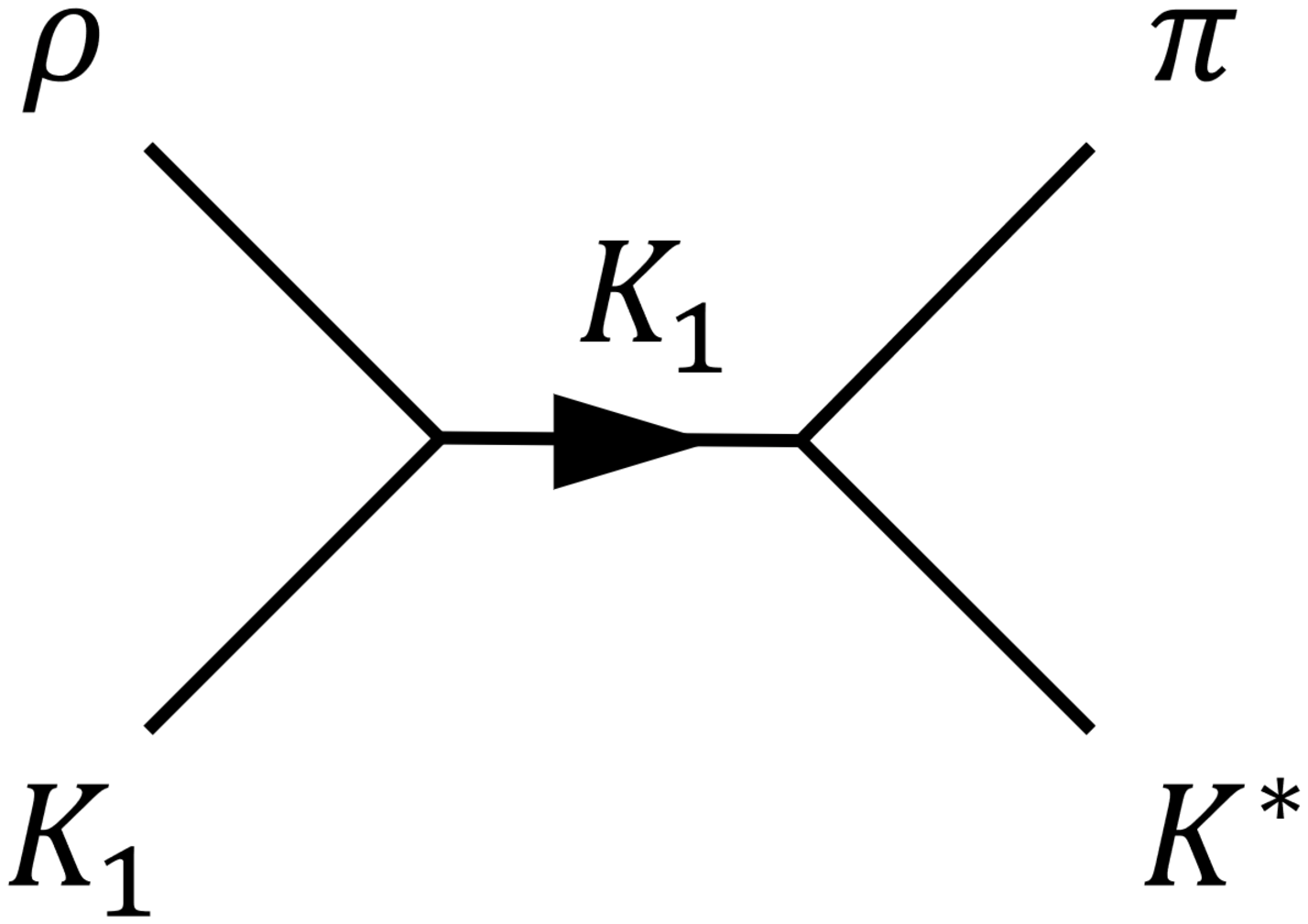}
\end{minipage}
\begin{minipage}[t]{0.24\textwidth}
    \includegraphics[height=2.3cm]{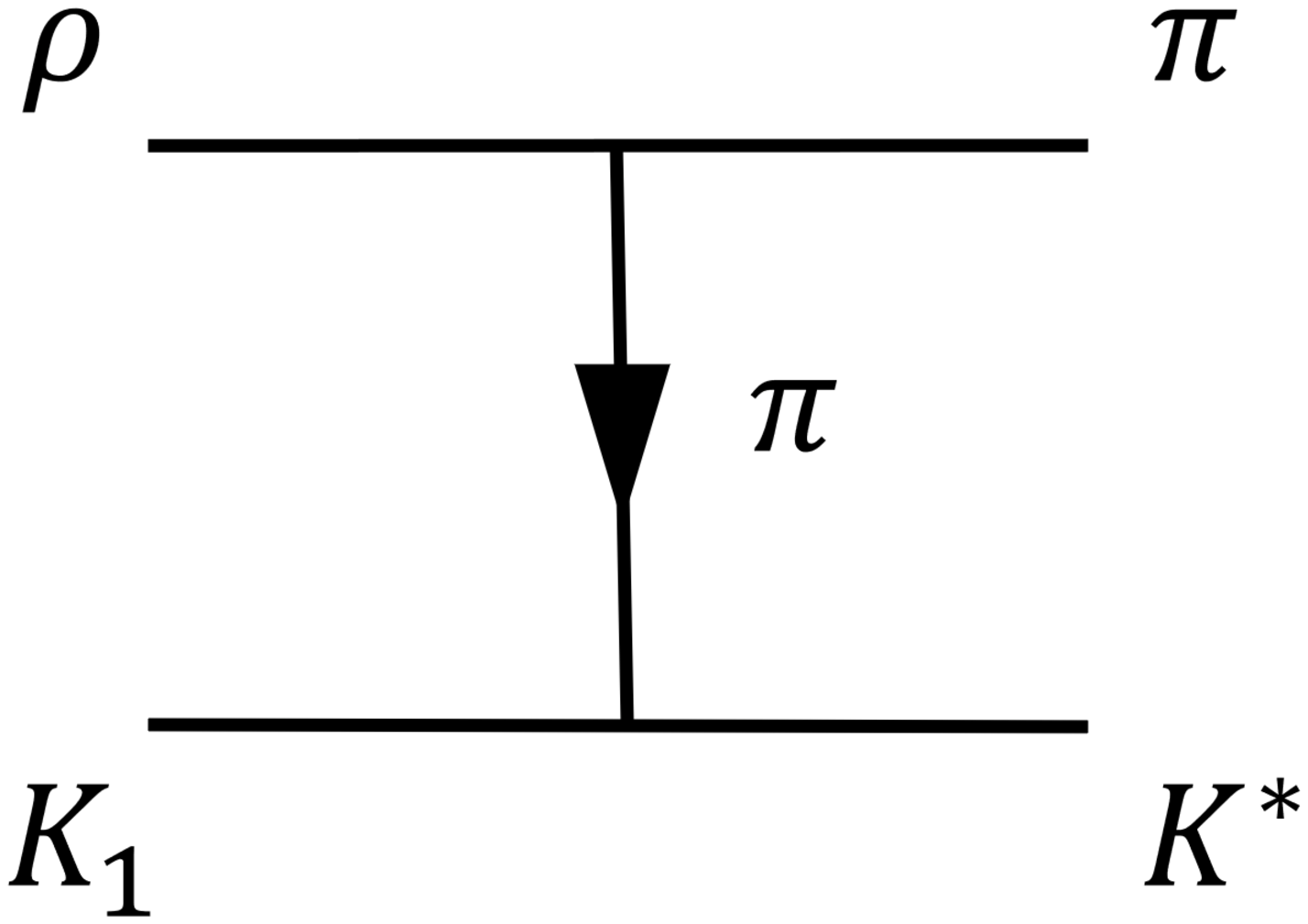}
\end{minipage}
\begin{minipage}[t]{0.24\textwidth}
    \includegraphics[height=2.3cm]{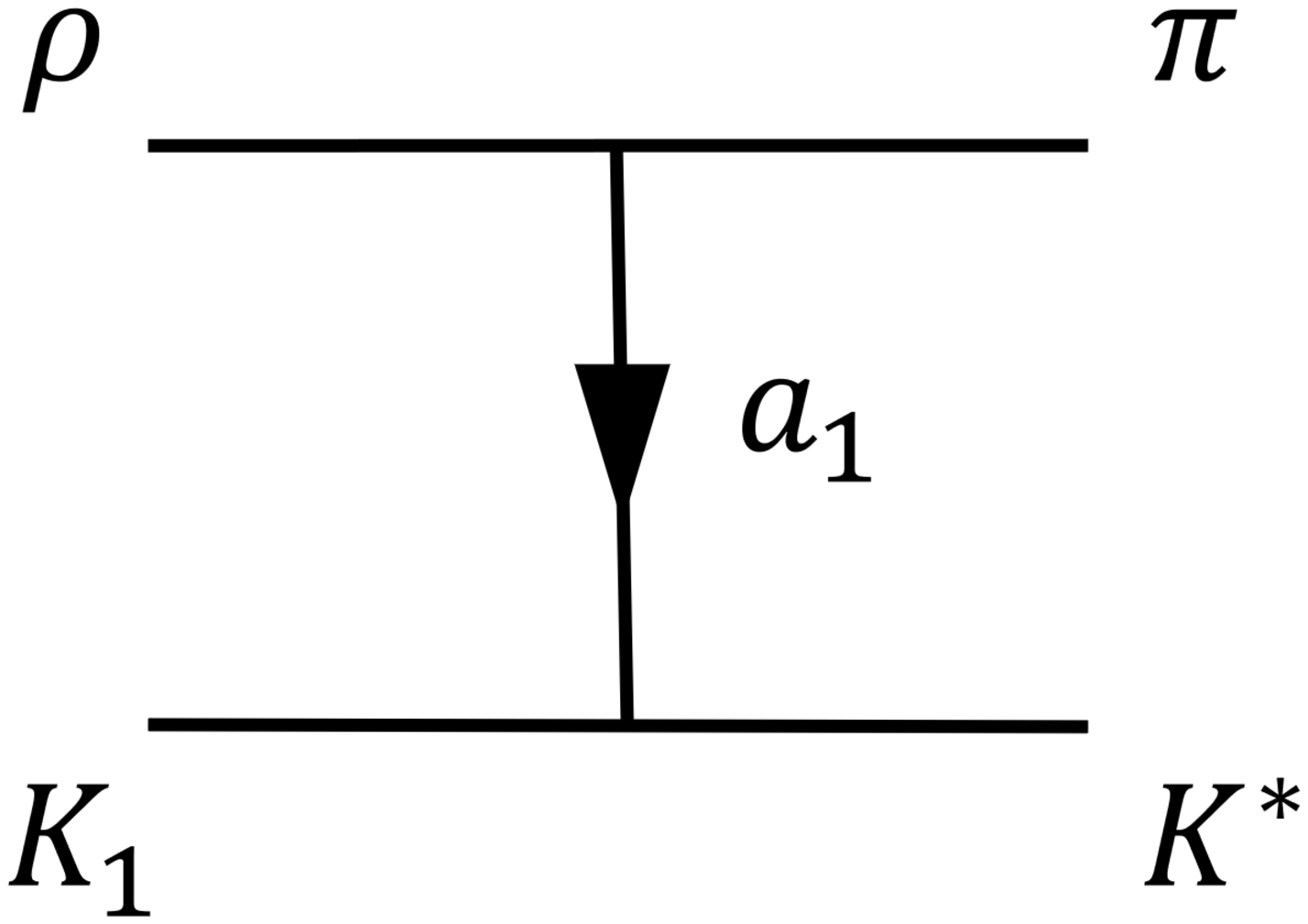}
\end{minipage}
\vspace{0.2cm}
\begin{minipage}{0.24\textwidth}
(3-a)
\end{minipage}
\begin{minipage}{0.252\textwidth}
(3-b)
\end{minipage}
\begin{minipage}{0.245\textwidth}
(3-c)
\end{minipage}
\begin{minipage}{0.24\textwidth}
(3-d)
\end{minipage}
\vspace{0.2cm}
\vspace{0.05cm}
\begin{minipage}[t]{0.18\textwidth} 
\includegraphics[height=1.9cm]{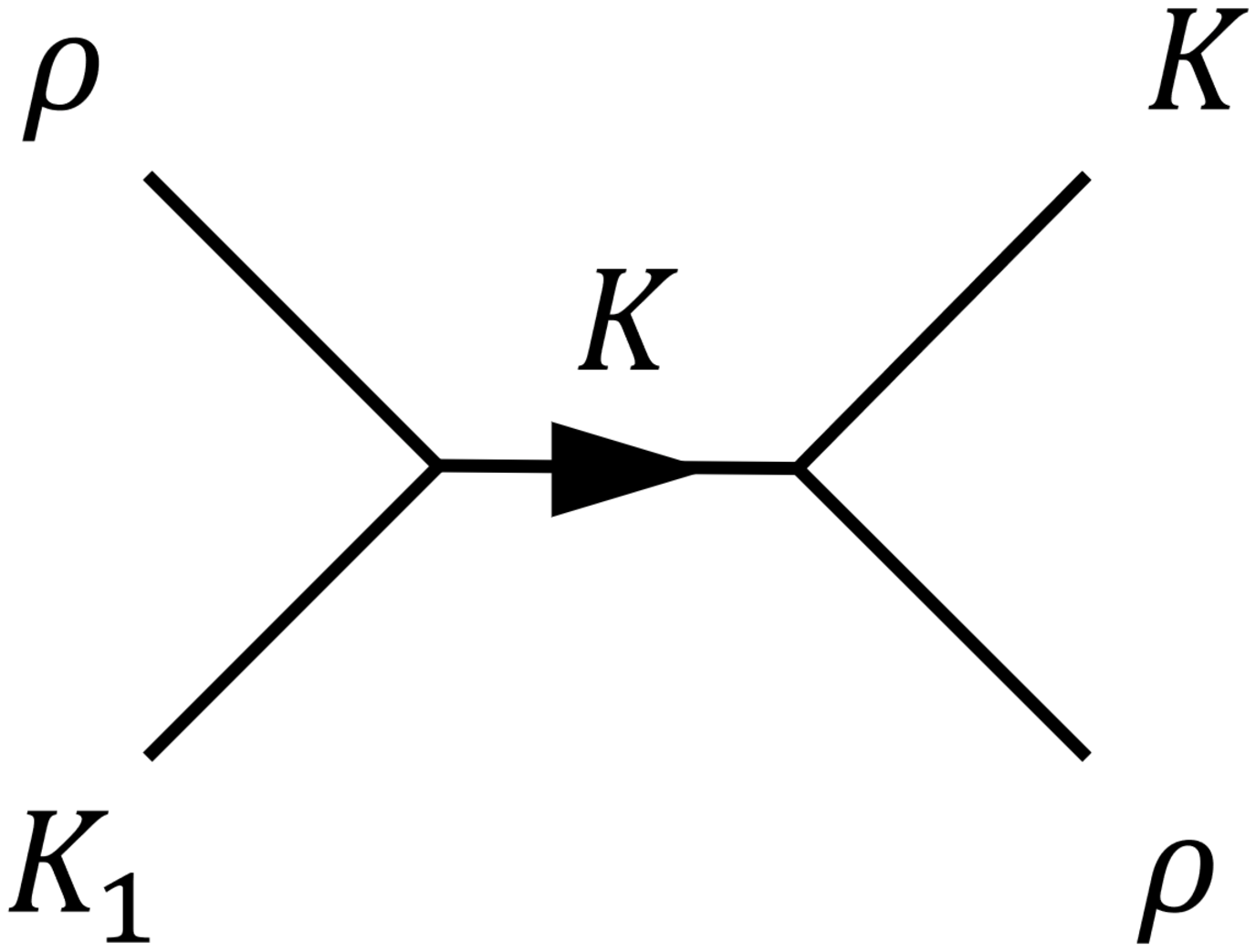}
\end{minipage}
\begin{minipage}[t]{0.18\textwidth}
        \includegraphics[height=1.9cm]{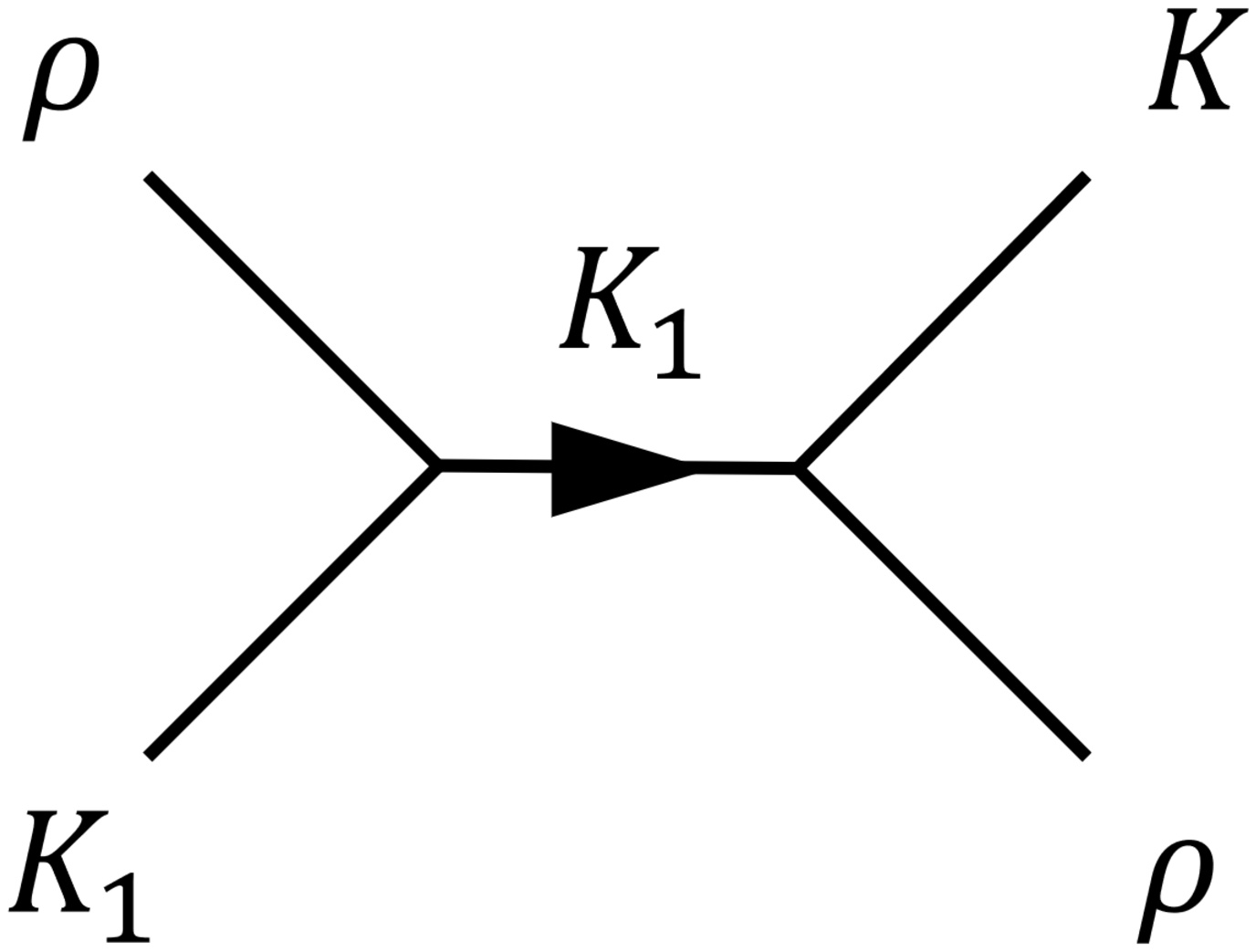}
\end{minipage}
\begin{minipage}[t]{0.18\textwidth}
        \includegraphics[height=1.9cm]{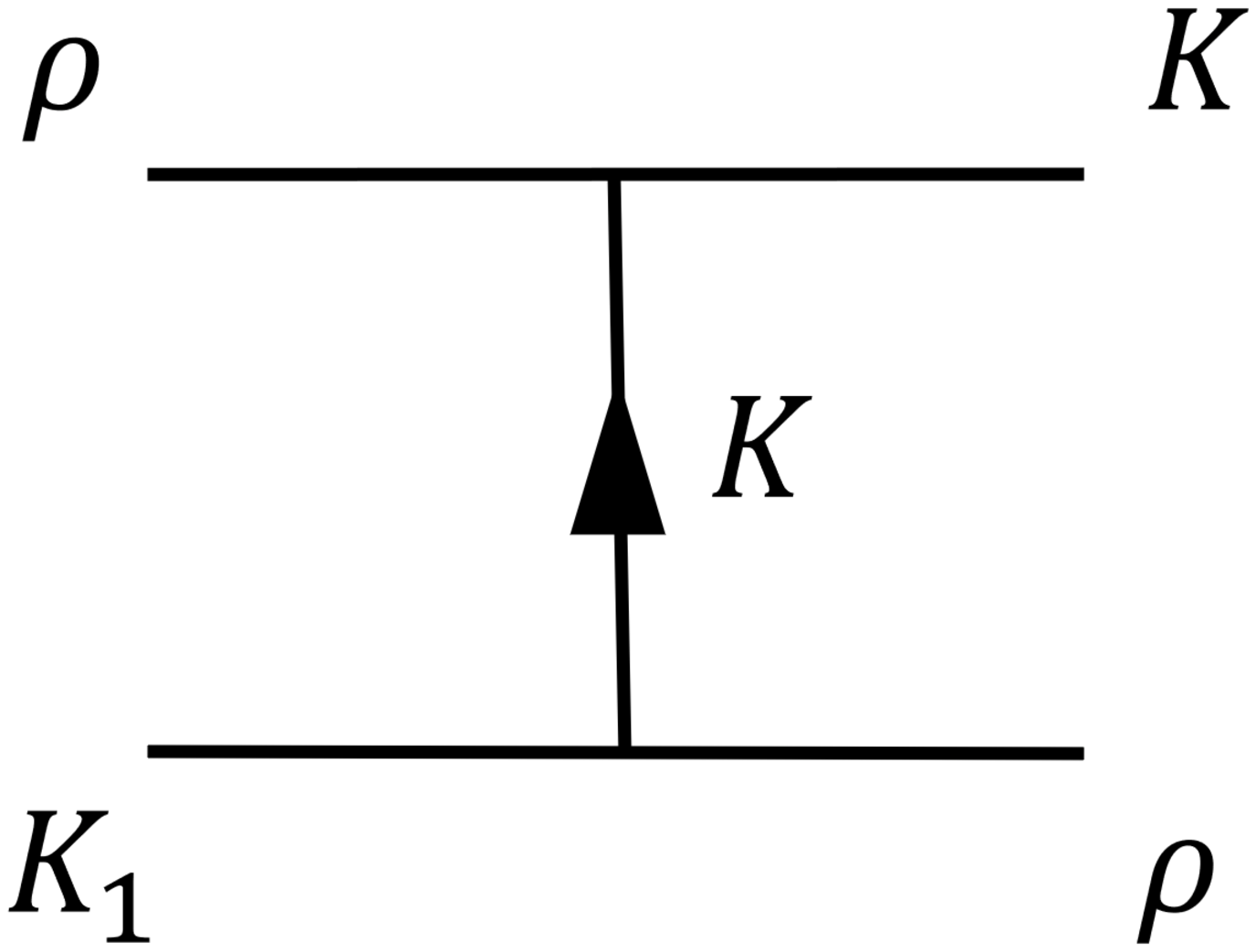}
\end{minipage}
\begin{minipage}[t]{0.18\textwidth}
 \includegraphics[height=1.9cm]{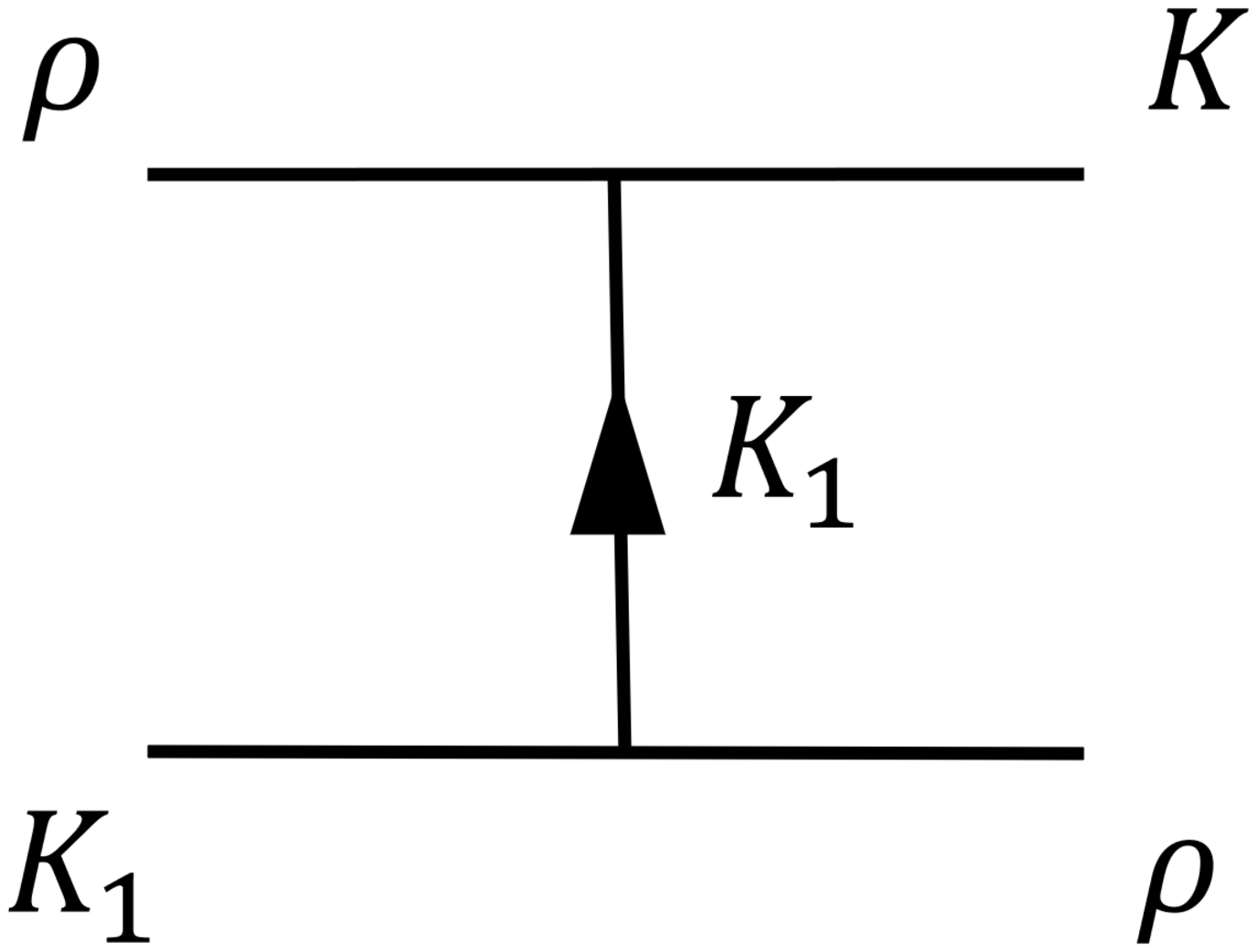}
\end{minipage}
\begin{minipage}[t]{0.19\textwidth} 
\includegraphics[height=1.9cm]{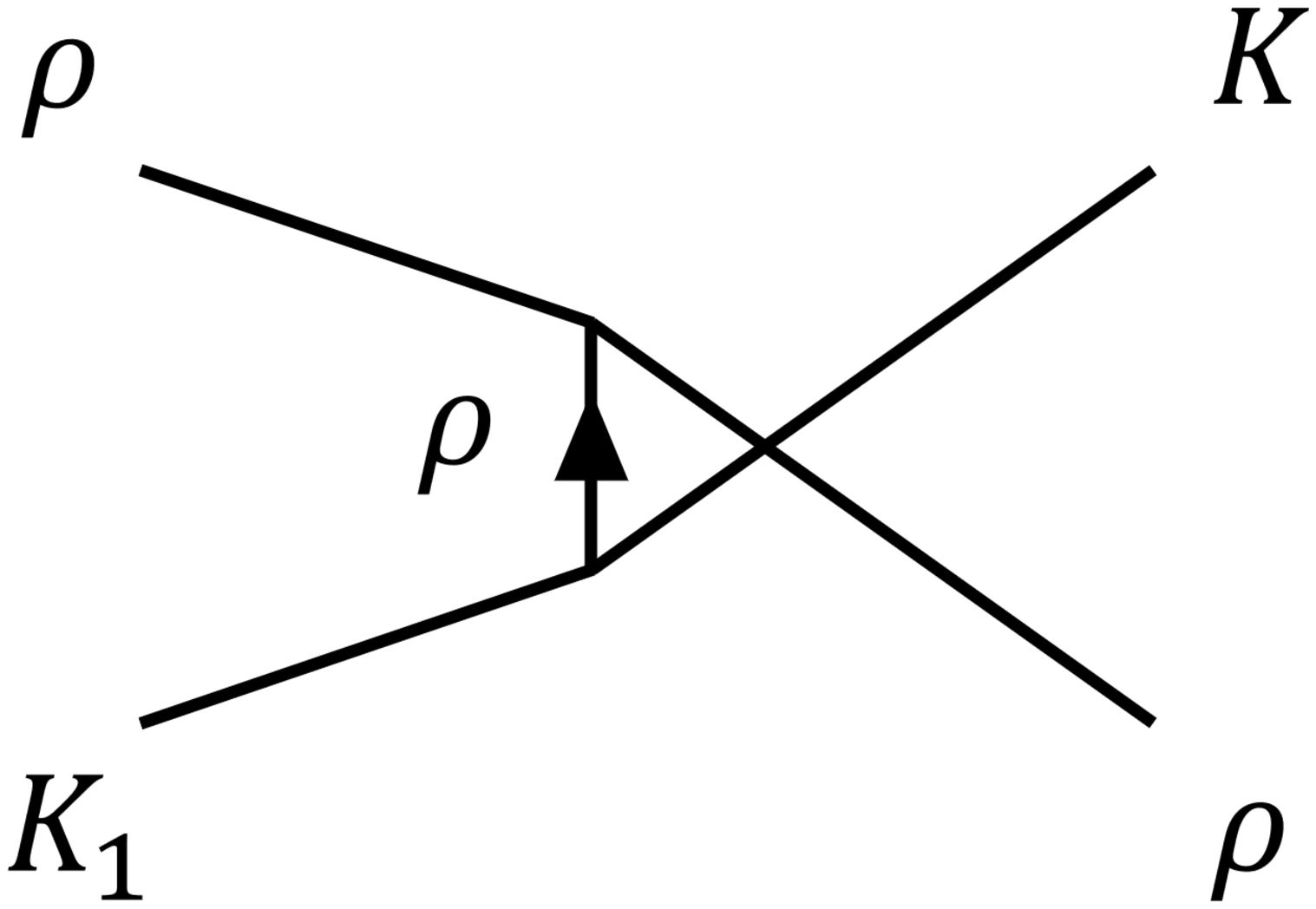}
\end{minipage}
\vspace{0.2cm}
\begin{minipage}{0.18\textwidth}
(4-a)
\end{minipage}
\begin{minipage}{0.18\textwidth}
(4-b)
\end{minipage}
\begin{minipage}{0.19\textwidth}
(4-c)
\end{minipage}
\begin{minipage}{0.18\textwidth}
(4-d)
\end{minipage}
\begin{minipage}{0.19\textwidth}
(4-e)
\end{minipage}
\vspace{0.1cm}
\caption{Feynman diagrams for the $K_{1}$ meson absorption by $\pi,\,\rho$ mesons. (1) $K_{1}\pi\rightarrow K\pi$, (2) $K_{1}\pi \rightarrow K^{*}\rho$, (3) $K_{1}\rho \rightarrow K^{*}\pi$ and (4) $K_{1}\rho \rightarrow K \rho$.
}
\label{fig:diagrams}
\end{figure*}


The amplitude for each scattering of different charge
states can be written as two isospin channels because of isospin
conservation. For example, the scattering of $\pi K_{1}
\rightarrow K \pi$ consists of the following charge states :
$K^{-}\pi^{-}\rightarrow K_{1}^{-}\pi^{-}$,
$K^{-}\pi^{0}\rightarrow K_{1}^{-}\pi^{0}$,
$\bar{K}^{0}\pi^{-}\rightarrow K_{1}^{-}\pi^{0}$, $K^{-}\pi^{+}
\rightarrow K_{1}^{-}\pi^{+}$, and $\bar{K}^{0}\pi^{0} \rightarrow
K_{1}^{-}\pi^{+}$. These five processes are described by two
independent scattering amplitudes $a_{3/2}$ and $a_{1/2}$ for
isospin $3/2$ and $1/2$ channels, respectively, so that
\begin{align}
\begin{cases}
 \langle K^{-}\pi^{-}|S|K_{1}^{-}\pi^{-}\rangle \,=\, a_{3/2} \cr
 \langle K^{-}\pi^{0}|S|K_{1}^{-}\pi^{0}\rangle \,=\, \frac{2}{3}a_{3/2}+\frac{1}{3}a_{1/2}
\cr
 \langle\bar{K}^{0}\pi^{-}|S|K_{1}^{-}\pi^{0}\rangle \,=\, -\frac{\sqrt{2}}{3}\left(a_{3/2}-a_{1/2}\right)
\cr
 \langle K^{-}\pi^{+}|S|K_{1}^{-}\pi^{+}\rangle \,=\, \frac{1}{3}a_{3/2}+\frac{2}{3}a_{1/2}
\cr
 \langle \bar{K}^{0}\pi^{0}|S|K_{1}^{-}\pi^{+}\rangle \,=\, \frac{\sqrt{2}}{3}\left(a_{3/2}-a_{1/2}\right)
\end{cases} .
\end{align}
It should be noted that squaring and adding all five amplitudes,
one finds that the sum is proportional to $2a^2_{3/2}+a^2_{1/2}$
reflecting the degeneracy of the isospin.

Using the interaction terms in the Lagrangian discussed above, we
evaluated the amplitudes for all diagrams shown in Fig.
\ref{fig:diagrams}. These represent the absorption amplitudes of
the $K_{1}$ meson by $\pi$ and $\rho$ mesons. First, the
amplitudes for $K_1$ absorption by $\pi$ mesons are represented by
the processes $\pi K_{1} \rightarrow \pi K$ and $\pi K_{1}
\rightarrow \rho K^{*}$. The amplitude for definite isospin
channels of the  $\pi(p_{1}) K_{1}(p_{2}) \rightarrow \pi(p_{3})
K(p_{4})$  are given as $a_{1/2} =
g_{K_{1}K\pi\pi}\epsilon_{2}^{\mu}(p_{1}+3p_{3})_{\mu}+3\mathcal{M}^{s}_{1}+2\mathcal{M}^{t}_{1}+\mathcal{M}^{u}_{1}$
and $a_{3/2} = -2
g_{K_{1}K\pi\pi}\epsilon_{2}^{\mu}p_{1\,\mu}-\mathcal{M}^{t}_{1}-2\mathcal{M}^{u}_{1}$
for isospin 1/2 and 3/2, respectively. Here, $\mathcal{M}^{s}_{1}
= \mathcal{M}_{(1-b)}$, $\mathcal{M}^{t}_{1} =
\mathcal{M}_{(1-c)}$,  and $\mathcal{M}^{u}_{1} =
\mathcal{M}_{(1-d)}$, where the subscript in the right hand side
represents the diagrams in Fig. \ref{fig:diagrams}.  The matrix
elements are given as
\begin{eqnarray}\label{eq:Amplitude_1}
\mathcal{M}^{s}_{1} &=& g_{K_{1}K^{*}\pi}g_{K^{*}K\pi}\, \epsilon_{2}^{\mu}\left(p_{3}-p_{4}\right)^{\nu}\nonumber\\
&&\times\frac{-g_{\mu\nu}+(p_{1}+p_{2})_{\mu}(p_{1}+p_{2})_{\nu}/s}{s-m_{K^{*}}^{2}} ,
\nonumber \\
\mathcal{M}^{t}_{1} &=&g_{\rho\pi\pi}g_{K_{1}K\rho}\, \epsilon_{2}^{\mu}\frac{-\left(p_{1}+p_{3 }\right)_{\mu}}{t-m_{\rho}^{2}} ,
\nonumber \\
\mathcal{M}^{u}_{1}&=&g_{K_{1}K^{*}\pi}g_{K^{*}K\pi}\, \epsilon_{2}^{\mu}\left(p_{1}+p_{4}\right)^{\nu}\nonumber\\
&&\times\frac{-g_{\mu\nu}+(p_{1}-p_{4})_{\mu}(p_{1}-p_{4})_{\nu}/u}{u-m_{K^{*}}^{2}}.
\nonumber \\
\end{eqnarray}
\begin{widetext}

The two independent amplitudes for the $\pi(p_{1}) K_{1}(p_{2})
\rightarrow \rho(p_{3}) K^{*}(p_{4})$ are respectively  given as $a_{1/2}
= 3\mathcal{M}^{s}_{2}+2\mathcal{M}^{t}_{2}+\mathcal{M}^{u}_{2}$
and $a_{3/2} = -\mathcal{M}^{t}_{2}-2\mathcal{M}^{u}_{2}$. Here,
$\mathcal{M}^{t}_{2} = \mathcal{M}_{(2-b)} +\mathcal{M}_{(2-d)}$
and $\mathcal{M}^{u}_{2} = \mathcal{M}_{(2-c)}
+\mathcal{M}_{(2-e)}$, with
\begin{eqnarray}\label{eq:Amplitude_2}
\mathcal{M}^{s}_{2} &=& g_{K_{1}K^{*}\pi}g_{K^{*}K^{*}\rho}\epsilon_{2 \,\alpha}\epsilon^{* \gamma}_{3}\epsilon^{*\delta}_{4}\frac{-g^{\alpha\beta}+(p_{1}+p_{2})^{\alpha}(p_{1}+p_{2})^{\beta}/s}{s-m_{K^{*}}^{2}}[g_{\gamma\beta}(2p_{3}+p_{4})_{\delta}-g_{\delta\beta}(p_{3}+2p_{4})_{\gamma}-g_{\gamma\delta}(p_{3}-p_{4})_{\beta}],
\nonumber \\
\mathcal{M}^{t}_{2} &=& g_{K_{1} K^{*}\pi}g_{\rho\pi\pi}\epsilon^{\mu}_{2}\epsilon^{*}_{4 \,\mu}\epsilon^{*\gamma}_{3}\frac{1}{t-m_{\pi}^{2}}(2p_{1}-p_{3})_{\gamma}
\nonumber \\
&&+ g_{a_{1}\rho\pi}g_{K_{1}a_{1}K^{*}}\epsilon^{\mu}_{2}\epsilon^{*}_{3\,\alpha}\epsilon^{*\,\delta}_{4}\frac{-g^{\alpha\beta}+(p_{1}-p_{3})^{\alpha}(p_{1}-p_{3})^{\beta}/t}{t-m_{a_{1}}^{2}}\left[g_{\mu\beta}p_{2\,\delta}-g_{\mu\delta}p_{2\,\beta}+g_{\beta\delta}(p_{1}-p_{3})_{\mu}\right],
\nonumber \\
\mathcal{M}^{u}_{2} &=& g_{K_{1} K\rho}g_{K^{*}K\pi}\epsilon^{\mu}_{2}\epsilon^{*}_{3\, \mu}\epsilon^{*\delta}_{4}\frac{1}{u-m_{\pi}^{2}}(2p_{1}-p_{4})_{\delta}
\nonumber \\
&&+g_{K_{1}K^{*}\pi}g_{K_{1}K_{1}\rho}\epsilon^{\mu}_{2}\epsilon^{*\,\gamma}_{3}\epsilon^{*}_{4\,\alpha}\frac{-g^{\alpha\beta}+(p_{1}-p_{4})^{\alpha}(p_{1}-p_{4})^{\beta}/u}{u-m_{K_{1}}^{2}}\left[g_{\mu\beta}(p_1-p_4)_{\gamma}+g_{\mu\gamma}p_{2\,\beta}-g_{\gamma\beta}(p_1-p_4)_{\mu}\right].
\nonumber \\
\end{eqnarray}

Second, the amplitudes for absorption by $\rho$ mesons are  $\rho K_{1}
\rightarrow \pi K^{*}$ and  $\rho K_{1} \rightarrow K \rho$.
The isospin channels of $\rho(p_{1}) K_{1}(p_{2}) \rightarrow
\pi(p_{3}) K^{*}(p_{4})$ are $a_{1/2} =
3\mathcal{M}^{s}_{3}-2\mathcal{M}^{t}_{3}$ and $a_{3/2} =
\mathcal{M}^{t}_{3}$. Here, $\mathcal{M}^{s}_{3} =
\mathcal{M}_{(3-a)} +\mathcal{M}_{(3-b)}$, and
$\mathcal{M}^{t}_{3} = \mathcal{M}_{(3-c)} +\mathcal{M}_{(3-f)}$
are given as follows:
\begin{eqnarray}\label{eq:Amplitude_3}
\mathcal{M}^{s}_{3} &=& g_{K_{1}K\rho}g_{K^{*}K\pi}\epsilon_{1}^{\mu}\epsilon_{2\,\mu}\epsilon_{4}^{*\,\lambda}\frac{1}{s-m_{K}^{2}}(2p_{3}+p_{4})_{\lambda}
\nonumber \\
&+&  g_{K_{1}K_{1}\rho}g_{K_{1}K\pi}\epsilon_{1}^{\mu}\epsilon_{2}^{\nu}\epsilon_{4}^{*\,\lambda}\frac{-g^{\mu '}_{\lambda}+(p_{1}+p_{2})^{\mu '}(p_{1}+p_{2})_{\lambda}/m_{K_{1}}^{2}}{s-m_{K_{1}}^{2}}[g_{\mu '\nu}(p_{1}+p_{2})_{\mu}-g_{\mu\mu '}(p_{1}+p_{2})_{\nu} + g_{\mu\nu}p_{2\,\mu '}],
\nonumber \\
\mathcal{M}^{t}_{3} &=& g_{K_{1}K^{*}\pi}g_{\rho\pi\pi}\epsilon_{1}^{\mu}\epsilon_{2}^{\nu}\epsilon_{4\,\nu}^{*}\frac{1}{t-m_{\pi}^{2}}(p_{1}-2p_{3})_{\mu}
\nonumber \\
&-&  g_{K_{1}a_{1}K^{*}}g_{a_{1}\rho\pi}\epsilon_{1}^{\mu}\epsilon_{2}^{\nu}\epsilon_{4}^{*\,\lambda}\frac{-g^{\nu '}_{\mu}+(p_{1}-p_{3})_{\mu}(p_{1}-p_{3})^{\nu '}/m_{a_{1}}^{2}}{t-m_{a_{1}}^{2}}[g_{\lambda\nu}p_{2\,\nu '}-g_{\nu\nu '}p_{2\,\lambda} - g_{\lambda\nu '}(p_{1}-p_{3})_{\nu}].
\end{eqnarray}
The isospin channels of  $\rho(p_{1}) K_{1}(p_{2}) \rightarrow
K(p_{3}) \rho(p_{4})$ are $a_{1/2} =
-3\mathcal{M}^{s}_{3}-\mathcal{M}^{t}_{4}+2\mathcal{M}^{u}_{4}$
and $a_{3/2} = 2\mathcal{M}^{t}_{4} -\mathcal{M}^{u}_{4}$. Here,
$\mathcal{M}^{s}_{4} = \mathcal{M}_{(4-a)} +\mathcal{M}_{(4-b)}$,
$\mathcal{M}^{t}_{4} = \mathcal{M}_{(4-c)} +\mathcal{M}_{(4-d)}$,
and $\mathcal{M}^{u}_{4} = \mathcal{M}_{(4-e)}$ are given as
follows:
\begin{eqnarray}\label{eq:Amplitude_4}
\mathcal{M}^{s}_{4} &=& g_{K_{1}K\rho}g_{\rho K K}\epsilon_{1}^{\mu}\epsilon_{2\,\mu}\epsilon_{4}^{*\,\lambda}\frac{1}{s-m_{K}^{2}}(2p_{3}+p_{4})_{\lambda}
\nonumber \\
&&+g_{K_{1}K_{1}\rho}g_{K_{1}K\rho }\epsilon_{1}^{\mu}\epsilon_{2}^{\nu}\epsilon_{4}^{*\,\lambda}\frac{-g^{\mu '}_{\lambda}+(p_{1}+p_{2})^{\mu '}(p_{1}+p_{2})_{\lambda}/m_{K_{1}}^{2}}{s-m_{K_{1}}^{2}}\left[g_{\nu\mu '}(p_{1}+p_{2})_{\mu}-g_{\mu\mu'}(p_{1}+p_{2})_{\nu}+g_{\mu\nu}p_{2\,\mu '}\right],
\nonumber \\
\mathcal{M}^{t}_{4} &=& g_{K_{1}K\rho}g_{\rho K K}\epsilon_{1}^{\mu}\epsilon_{2}^{\nu}\epsilon_{4\,\nu}^{*}\frac{1}{t-m_{K}^{2}}(p_{1}-2p_{3})_{\mu}
\nonumber \\
&&+ g_{K_{1}K_{1}\rho}g_{K_{1}K\rho }\epsilon_{1}^{\mu}\epsilon_{2}^{\nu}\epsilon_{4}^{*\,\lambda}\frac{-g^{\nu '}_{\mu}+(p_{1}-p_{3})_{\mu }(p_{1}-p_{3})^{\nu '}/m_{K_{1}}^{2}}{t-m_{K_{1}}^{2}}\left[g_{\nu\nu '}(p_{1}-p_{3})_{\lambda}-g_{\nu ' \lambda}(p_{1}-p_{3})_{\nu}+g_{\lambda\nu}p_{2\,\nu '}\right],
\nonumber \\
\mathcal{M}^{u}_{4} &=& g_{K_{1}K\rho}g_{\rho\rho\rho}\epsilon_{1}^{\mu}\epsilon_{2}^{\nu}\epsilon_{4}^{*\,\lambda}\frac{-g^{\mu '}_{\nu}+(p_{1}-p_{4})^{\mu '}(p_{1}-p_{4})_{\nu}/m_{\rho}^{2}}{u-m_{\rho}^{2}}
\left[g_{\mu\mu '}\left(2p_{1}-p_{4}\right)_{\lambda}-g_{\mu '\lambda}(p_{1}-2p_{4})_{\mu}-g_{\mu\lambda}(p_{1}+p_{4})_{\mu '}\right].
\nonumber \\
\end{eqnarray}
\end{widetext}
We keep the convention where particles 1 and 2 stand for
initial-state mesons, and particles 3 and 4 stand for final-state
mesons shown on the left and right sides of the diagrams,
respectively. The Mandelstam variables $s=(p_{1}+p_{2})^{2}$,
$t=(p_{1}-p_{3})^{2}$, and $u=(p_{1}-p_{4})^{2}$ have also been
used.

To take into account the finite size of the hadrons when
calculating amplitudes, we introduce form factors shown below at
each interaction vertex for the $u,t-$channel and the
$s-$channel, respectively,
\begin{eqnarray}\label{eq:Form factor_stu}
F_{u,t}(\vec{q}) = \frac{\Lambda^{2}-m_{ex}^{2}}{\Lambda^{2}+\vec{q}^{2}},
\qquad
 F_{s}(\vec{q}) = \frac{\Lambda^{2}+m_{ex}^{2}}{\Lambda^{2}+\omega^{2}},
\end{eqnarray}
with $m_{ex}$ being the mass of the exchanging particle at each
diagram. In Eq. (\ref{eq:Form factor_stu}) $\vec{q}^{2}$ is the
squared three-momentum transfer at $t,u-$channels, and
$\omega^{2}$ is the total energy of the incoming particles at the
$s-$channel in the center of mass frame. On the other hand, we
apply the following form factor for the four point contact
interaction,
\begin{eqnarray}\label{eq:Form factor_c}
F_{c}(\vec{k}) = \left(\frac{\Lambda^{2}}{\Lambda^{2}+\vec{k}^{2}}\right)^{2},
\end{eqnarray}
where $\vec{k}$ is the average of the squared three-momenta from
the form factors for the given channels at each process. We set
the cut off parameter $\Lambda=$ 1.8 GeV \cite{Brown:1991ig}.

\begin{figure}[!t]
\begin{center}
\includegraphics[width=0.45\textwidth]{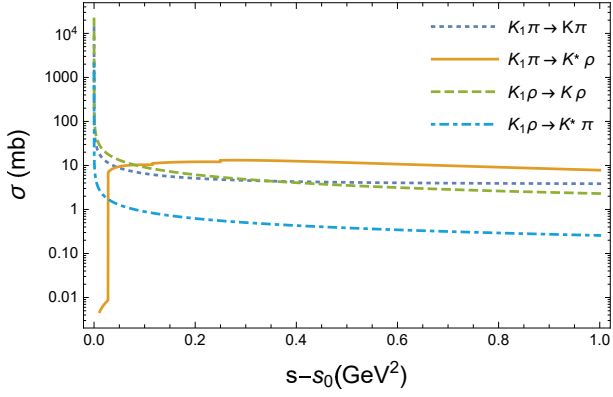}
\end{center}
\caption{The cross sections for the dissociation of the $K_{1}$ meson by $\pi$ and $\rho$ mesons via processes
$K_{1}\pi\rightarrow K\pi$,
$K_{1}\pi\rightarrow K^{*}\rho$,
$K_{1}\rho\rightarrow K\rho$ and $K_{1}\rho \rightarrow K^{*}\pi$.} \label{fig:Crs}
\end{figure}

The cross section after spin averaging is given by
\begin{eqnarray}
\sigma = \frac{1}{64\pi s}\int dt\frac{\overline{\mathcal{M}}^{2}}{|p_{1cm}|^{2}}F^{4}, \label{cross-section}
\end{eqnarray}
with $\overline{\mathcal{M}}^{2}$ being the squared amplitude of
all processes evaluated by averaging and summing over the
degeneracies of the initial and final particles, respectively.
$p_{1cm}$ in Eq. (\ref{cross-section}) represents for the three
momenta of the initial particles in the center of mass frame. Fig.
\ref{fig:Crs} shows the cross section for the absorption of the
$K_{1}$ meson by $\pi$ and $\rho$ mesons.

For the hadronic production cross section of the $K_{1}$ meson
during the hadronic stage in heavy-ion collisions, we will use
detailed balance relation in the rate equation.

\section{Time evolution of the $K_{1}$ meson abundance}

We now investigate the time evolution of the $K_{1}$ meson
abundance using the cross sections evaluated in the previous
section. We construct an evolution equation, similar to that in
Ref. \cite{Cho:2015qca}, composed of densities and abundances for
mesons participating in all the processes shown in Fig.
\ref{fig:Crs}: the dissociation processes of $K_1$ are due to
$\pi$ and $\rho$ mesons.

\begin{eqnarray}\label{eq:rate}
\frac{dN_{K_{1}}(\tau)}{d\tau} & = &  \left< \sigma_{K_{1}\pi
\rightarrow K\pi}\,v_{K_{1}\pi} \right>
 n_{\pi} \bigg[ - N_{K_{1}} + \frac{N_{K_{1}}^T}{N_{K}^T}            N_{K} \bigg]
 \nonumber \\
&+& \left<\sigma_{K_{1}\pi \rightarrow K^{*} \rho}\,v_{K_{1}\pi}
\right> \bigg[-
 n_{\pi} N_{K_{1}} + \frac{ n_{\pi}^T N_{K_{1}}^T }{n_{\rho}^T N_{K^{*}}^T} n_{\rho} N_{K^{*}} \bigg]
 \nonumber \\
&+& \left< \sigma_{K_{1}\rho \rightarrow K^{*}\pi}\,v_{K_{1}\rho}
\right> \bigg[- n_{\rho}N_{K_{1}} + \frac{n_{\rho}^TN_{K_{1}}^T}{
n_{\pi}^T N_{K^{*}}^T} n_{\pi} N_{K^{*}} \bigg]
\nonumber \\ 
 &+&\left<\sigma_{K_{1}\rho \rightarrow K \rho}\,v_{K_{1}\rho} \right>
\bigg[ - n_{\rho}N_{K_{1}} +  \frac{ n_{\rho}^TN_{K_{1}}^T}{
n_{\rho}^T N_{K}^T}   n_{\rho} N_{K} \bigg]
 \nonumber \\
 &+& \left< \Gamma_{K_{1} \rightarrow K \rho} \right> \bigg[-
N_{K_{1}} + \frac{ N_{K_{1}}^T}{  n_\rho^TN_{K}^T} n_{\rho}N_{K}
\bigg]
\nonumber \\
 &+& \left< \Gamma_{K_{1} \rightarrow K^* \pi} \right> \bigg[- N_{K_{1}} + \frac{N_{K_{1}}^T }{ n_{\pi}^TN_{K^{*}}^T } n_{\pi}N_{K^{*}} \bigg],
\end{eqnarray}
with $n_{i}(\tau)$ being the density of the light meson $i$ in the
hadronic phase at proper time $\tau$.
 $N_{j}(\tau) =
n_{j}(\tau)V(\tau)$, where $V(\tau)$ is the volume of the hadronic
matter given in Eq. \eqref{eq:TandV}. The ones with the
superscript $T$ are the thermal abundance at temperature $T$,
which depends on the proper time $\tau$ as given in Eq.
\eqref{eq:TandV}.
\begin{eqnarray}\label{eq:density}
n_{i}^T(\tau) &=& \frac{g_{i}}{2\pi^{2}}\int^{\infty}_{0}\frac{p^{2}dp}{\exp\left(\sqrt{p^{2}+m_{i}^{2}}/T(\tau)\right)-1}
\nonumber \\
&\approx & \frac{g_{i}}{2\pi^{2}}m_{i}^{2}T(\tau)K_{2}\left[ \frac{m_{i}}{T(\tau)}\right], 
\end{eqnarray}
where $K_{2}$ is the modified Bessel function of the second kind.
We take $n_{i}(\tau)=n_{i}^T(\tau)$ for pions and $\rho$ mesons.

We use the following two constraint equations involving the $\tau$
dependence for the numbers of $K$ and $K^*$.
\begin{eqnarray}
N_{K_1}+N_{N_{K^*}}+N_K & = &  N_0, \label{constant} \\
\frac{N_{K^*}+N_{K_1}\times {\rm BR} }{N_0} & = & aT+b, \label{ksk}
\end{eqnarray}
where $N_0$  is a constant determined at the chemical freeze-out
point.  Eq.~(\ref{constant}) follows from neglecting strangeness
annihilation  during the hadronic phase.  BR=0.16 is the branching
ratio of $K_1$ decaying into $K^*$ so that the ratio in Eq.
(\ref{ksk}) is the ratio between the final $K^*$ and $K$ numbers
when these mesons freeze-out at temperature $T$. This equation  is
motivated by the experimental observation that show the
approximately linear decrease in the observed ratio between $K^*$
and $K$ numbers as a function of the cube root of multiplicity in
heavy ion collision~\cite{Acharya:2019qge}, which can be
understood in a hadronic model calculation \cite{Cho:2015qca}
that finds the ratio to decreases linearly with the temperature of
the hadronic phase. The coefficients $a$ and b in Eq.~(\ref{ksk})
were determined to reproduce the statistical model prediction for
the ratio at $T=T_c$ and its linear reduction to 0.158 at $T=90$
MeV, which are the observed values for the ratio and the extracted
freeze-out temperature~\cite{Acharya:2019yoi} for  the lowest and
highest centrality   Pb-Pb collision~\cite{Acharya:2019qge},
respectively. $a=$ 1.816 GeV$^{-1}$ and $b=-$0.005444 are obtained
for the first scenario, where we assume that the number of $K_1$
is equal to that of $K^*$ at the chemical freeze-out point,
whereas $a=$ 2.272 GeV$^{-1}$ and $b=-$0.04643 in the second
scenario, where we assume that both yields of $K_1$ and $K^*$
follow statistical model values at the chemical freeze-out point.
Solving Eq.~(\ref{eq:rate}) with the two constraints given in
Eq.~(\ref{constant}) and Eq.~(\ref{ksk}), we can determine the
$\tau$ dependence and the final freeze-out numbers of
$N_{K_1},N_{K^*},N_K$.

When substituting in the rate equation, we take the degeneracy
factor  $g_{i}=(2S+1)$, where $S$ is the spin, as the isospin
effect is taken into account in the cross section which includes
the contribution from all charge states. The production
contribution has been taken into account by a detailed balance
condition, which requires each square bracket in
Eq.~(\ref{eq:rate}) to be zero in thermal equilibrium.

In Eq. \eqref{eq:rate}, we have considered the thermally averaged
cross section, $\left<\sigma_{ab\rightarrow cd}v_{ab}\right>$
given below for initial two particles in a two-body process
$ab\rightarrow cd$,
\begin{eqnarray}\label{eq:AvgCrs}
\left<\sigma_{ab\rightarrow cd}v_{ab}\right> &=& \frac{\int d^{3}\vec{p}_{a}d^{3}\vec{p}_{b}f_{a}(\vec{p}_{a})f_{b}(\vec{p}_{b})\sigma_{ab\rightarrow cd}v_{ab}}{\int d^{3}\vec{p}_{a}d^{3}\vec{p}_{b}f_{a}(\vec{p}_{a})f_{b}(\vec{p}_{b})}
\nonumber \\
&=& \left[64\pi \,T m_{a}^{2} m_{b}^{2} K_{2}(m_{a}/T)K_{2}(m_{b}/T)\right]^{-1}
\nonumber \\
&&\times \int^{\infty}_{\sqrt{s_{0}}}d\sqrt{s}\,K_{1}(\sqrt{s}/T)\int_{t_{0}}^{t_{1}} dt\,|\mathcal{M}|^{2},
\end{eqnarray}
with $\sqrt{s_{0}} ={\rm max}(m_{a}+m_{b}, m_{c}+m_{d})$, $K_{1}$ and
$K_{2}$ being the modified Bessel function of the second kind. The integration limits are $t_{0}(t_{1})$ = $(\frac{m_{1}^2+m_{3}^2-m_{2}^2+m_{4}^2}{2\sqrt{s}})^2-(p_{1cm}\mp p_{3cm})^2$.
Here, we approximate $f_{i}$ to be the Boltzmann momentum
distribution of a particle $i$, $f_{i}(\vec{p}) =
e^{-\sqrt{\vec{p}^{2}+m^{2}}/T}$. $v_{ab}$ is the relative
velocity of particles of species $a$ and $b$, $v_{ab} =
\sqrt{\left(p_{a}\cdot
p_{b}\right)^{2}-m_{a}^{2}m_{b}^{2}}/(E_{a}E_{b})$.
$\left<\Gamma_{K_{1}}\right>=\Gamma_{K_{1}}(m_{K_1})K_1(m_{K_1}/T)/K_2(m_{K_1}/T)$
is the thermally averaged decay width of the $K_{1}$
meson~\cite{Cho:2015qca}.
\begin{figure}[!t]
\begin{center}
\includegraphics[width=0.45\textwidth]{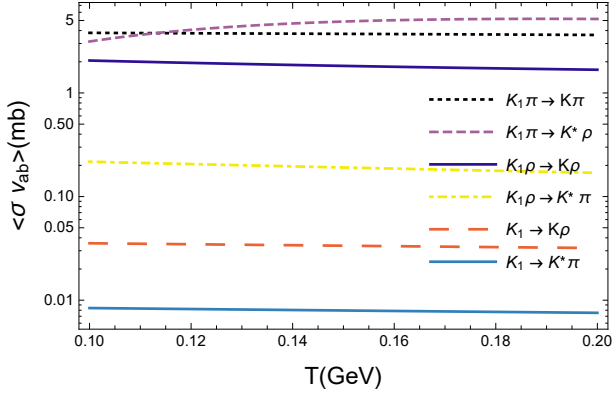}
\end{center}
\caption{Thermally averaged cross sections for the absorption of a $K_{1}$ meson via processes $K_{1}\pi \rightarrow K\pi$, $K_{1}\pi \rightarrow K^{*}\rho$, $K_{1}\rho \rightarrow K\rho$, $K_{1}\rho \rightarrow K^{*} \pi$, $K_{1}\rightarrow K \rho$, and $K_{1} \rightarrow K^{*}\pi$.}
\end{figure}

\section{initial temperature and time evolution in heavy ion collision}

The initial temperature inside the nuclear matter produced
from heavy ion collisions is studied with hydro simulation. We
will consider Pb-Pb collision at $\sqrt{s_{NN}}=5.02$ TeV in
different centrality ranges: 0--5 \%, 40--50 \% and 70--80 \%. The
initial energy density in the transverse plane is calculated by
using a two-dimensional Gaussian distribution of
$\sigma=0.4~\mathrm{fm}$ for each participating nucleon obtained
from the Monte Carlo Glauber framework~\cite{Miller:2007ri}. A
scale factor as a function of the number of participating nucleons
is multiplied to the initial density for a hydrodynamic evolution
with the \textsc{sonic} model~\cite{Habich:2014jna}. This tune is
required to match the multiplicity with the measured data in
various centrality ranges~\cite{Adam:2016ddh}.

\begin{figure}[!t]
\begin{center}
\includegraphics[width=0.4\textwidth]{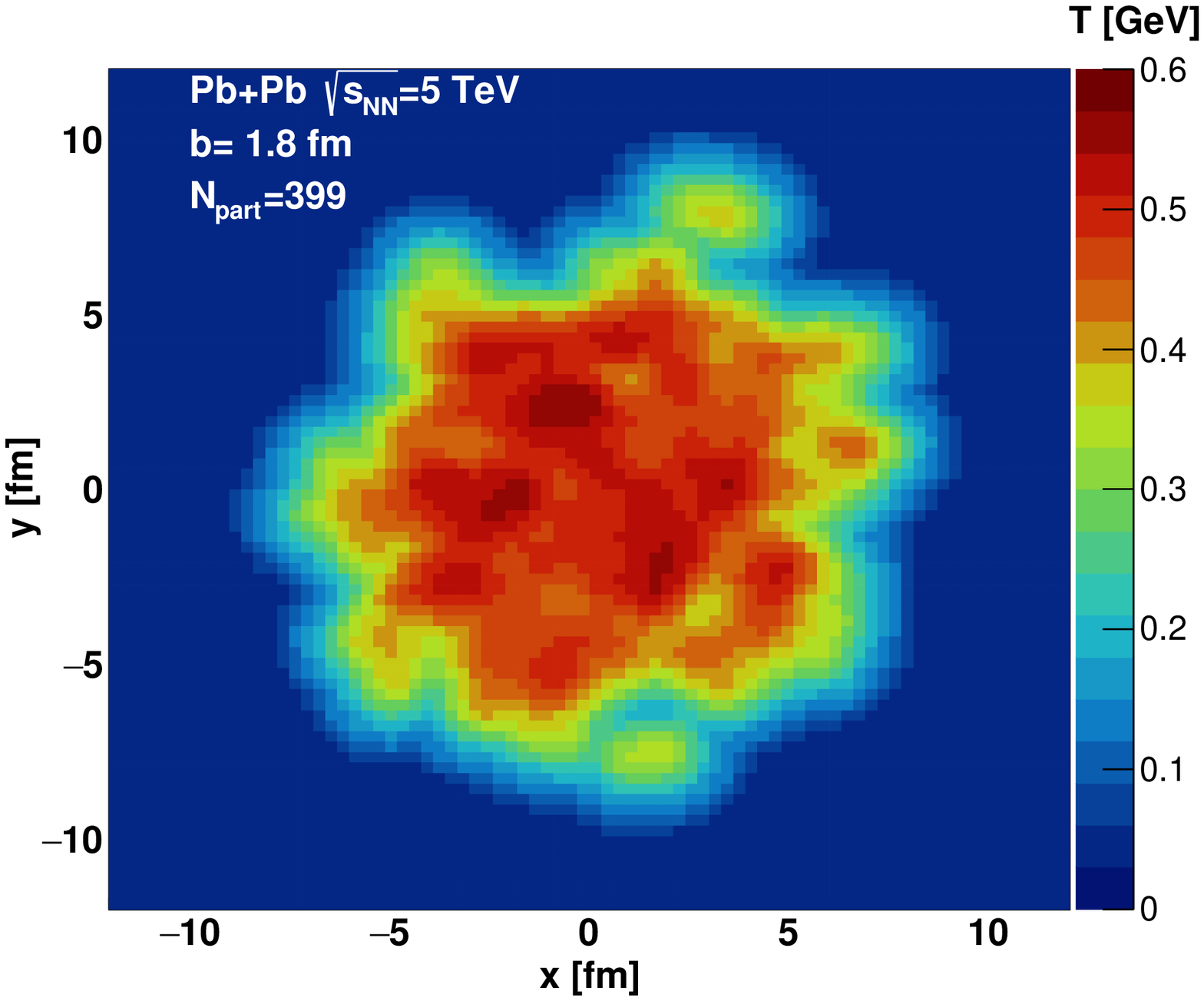}
\includegraphics[width=0.4\textwidth]{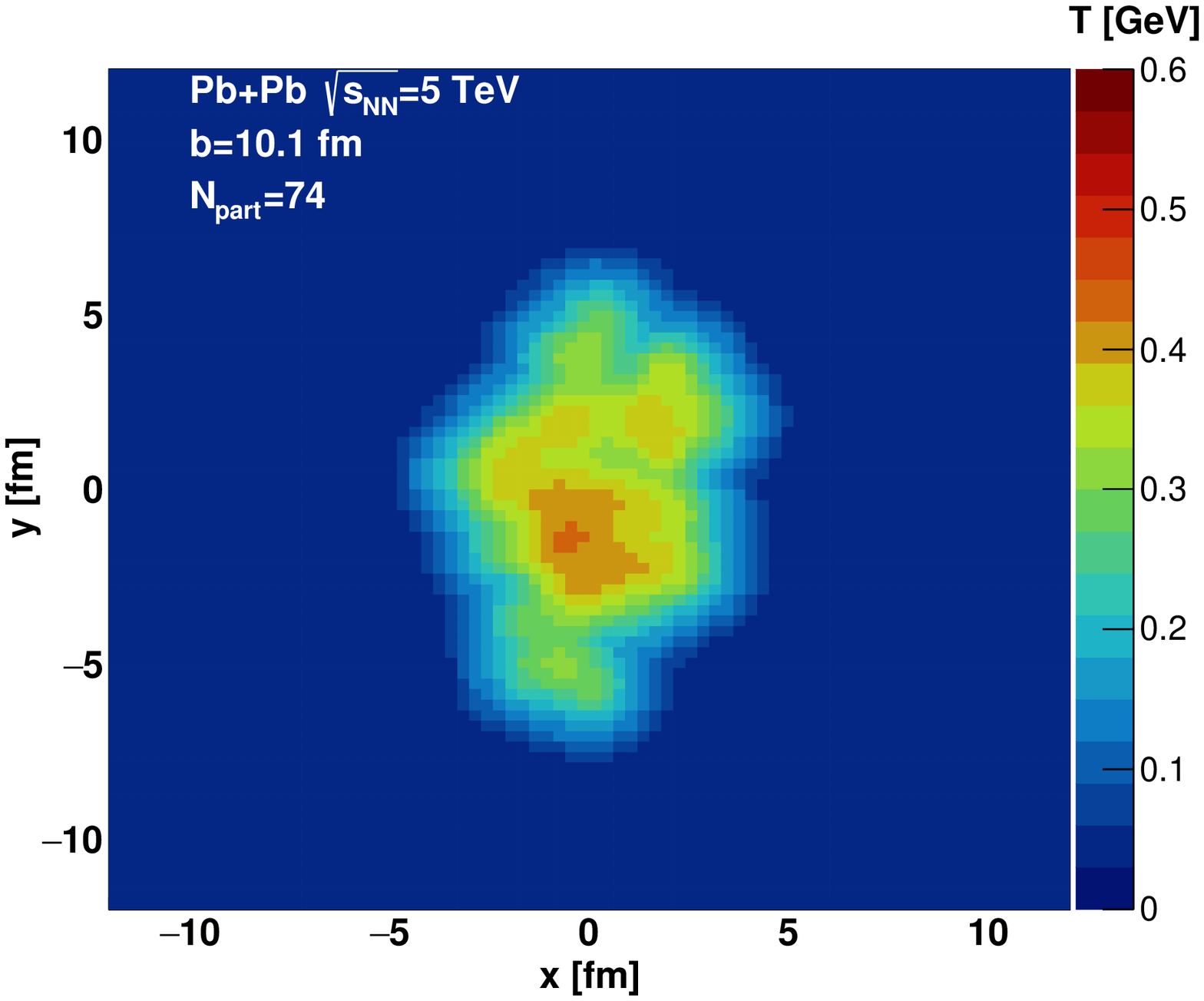}
\includegraphics[width=0.4\textwidth]{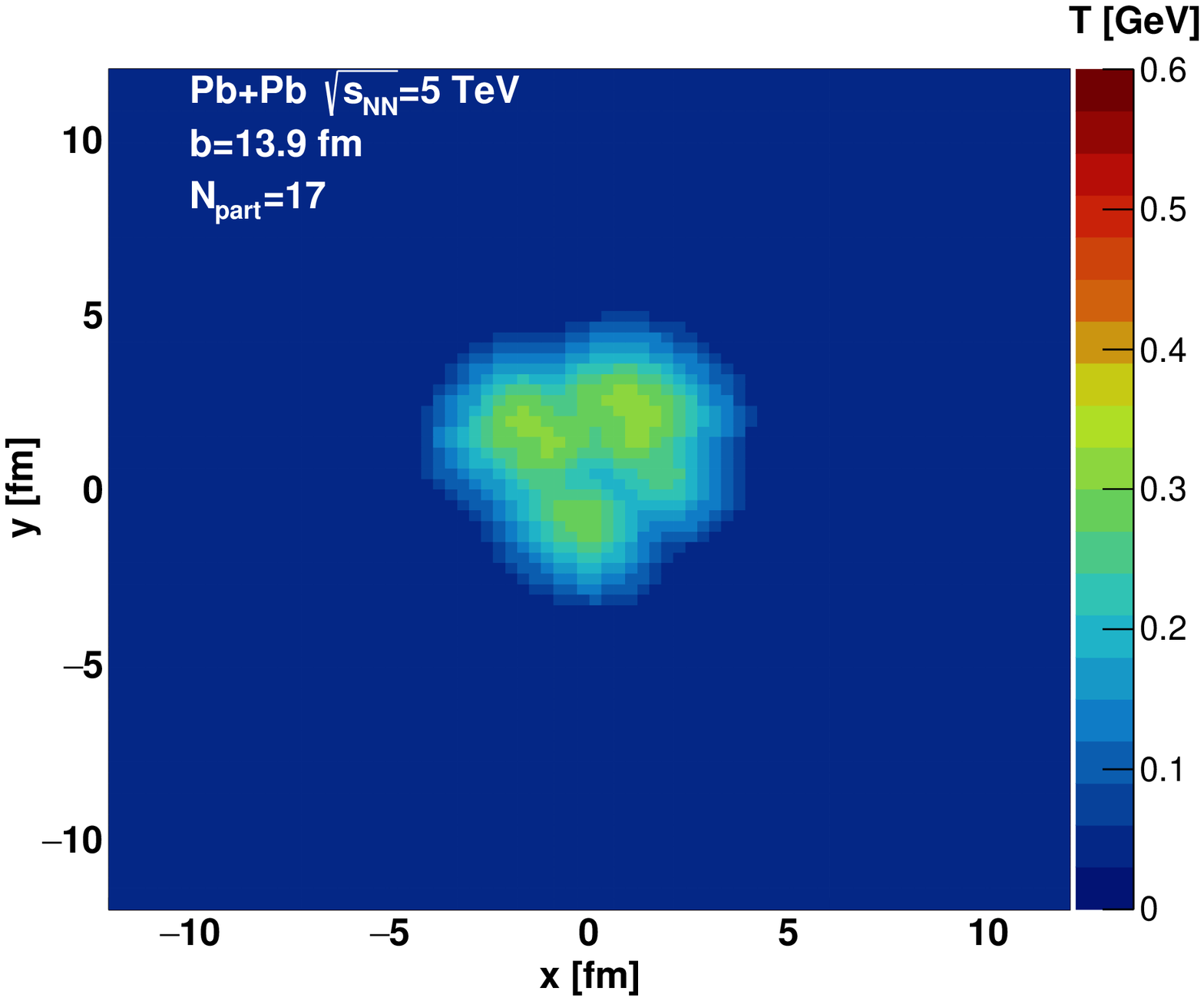}
\end{center}
 \caption{Initial temperature distribution for example events in 0--5\%, 40--50\% and 70--80\% centrality ranges of Pb-Pb collisions at $\sqrt{s_{NN}}=5.02$ TeV. }
  \label{fig:profile}
\end{figure}

Figure~\ref{fig:profile} shows the initial temperature distribution at
$\tau_{0}=0.5~{\rm fm}/c$ of example events in three centrality
ranges from the hydrodynamic simulation. As can be seen in the
figure, the fraction of area where initial temperature is higher
than 156 MeV is 84 \%, 75 \%, and 67 \% in the 0--5 \%, 40--50 \%,
and 70--80 \% centrality ranges, respectively. Furthermore, the
fraction of initial energy that comes from regions where the
temperature is higher than 156 MeV is larger than 98\% in all
three centrality ranges. Therefore we can assume that almost all
$K_{1}$ and $K^{*}$ are produced through the QGP phase even in  70--80
\% central collisions.

For calculation of the number of $K_{1}$ and $K^{*}$ mesons inside
the expanding medium, a more simple approach is used by assuming
an isentropic expansion of uniform matter as
follows~\cite{Song:2010fk}:

\begin{eqnarray}
&&\partial_\tau (A\tau \langle T^{\tau \tau}\rangle)=-pA,\label{energy7}\\
\nonumber\\
&&\partial_\tau (A\tau s \langle \gamma_r\rangle)=0,\label{entropy7}
\end{eqnarray}
where $T^{\tau \tau}=(e+p)u_\tau^2 -p$ is the energy momentum tensor in
the Milne coordinate system with $u_\tau$, $e$ and $p$ being the fluid
velocity, energy density and pressure, respectively. $s$ is the
entropy density, $\gamma_r=1/\sqrt{1-v_r^2}$
with $v_r$ being the radial velocity of fluid cell, $A=\pi R^2$ with
$R$ being the transverse radius of the uniform matter and
$\langle\cdots\rangle$ denotes average over the transverse area.
Eq.~(\ref{entropy7}) clearly shows that the total entropy is conserved
during the expansion with Lorentz contraction of transverse area
due to the flow velocity taken into account.

It should be noted that while the entropy conservation is a reasonable approximation for the QGP phase, the effect of larger viscosity has to be taken into account for the hadronic phase\cite{Demir:2008tr}.  
While the simple hydrodynamic model could be supplemented through  hadronic transport models to take into account the increase of entropy during the hadronic phase\cite{Xu:2017akx}, we will leave such improvements for later study and neglect such effects in this work as the main tendency of the enhancement as a function of centrality will remain valid.

Assuming that the radial flow velocity is a linear function of the
radial distance from the center, i.e., $\gamma_r v_r=\gamma_R
\dot{R}(r/R)$, where $\dot{R}=\partial R/\partial \tau$ and
$\gamma_R=1/\sqrt{1-\dot{R}^2}$, we then have~\cite{Song:2010fk}
\begin{eqnarray}
&&\langle u_\tau^2\rangle|_{\eta=0}=\langle\gamma_r^2\rangle=1+\frac{\gamma_R^2 \dot{R}^2}{2},\nonumber\\
&&\langle\gamma_r\rangle=\frac{2}{3\gamma_R^2 \dot{R}^2}\left(\gamma_R^3-1\right).
\label{gamma}
\end{eqnarray}
Since the energy density $e$ and pressure $p$ are related by the
equation of state of the matter through its temperature
$T$~\cite{Borsanyi:2010cj}, Eqs.(\ref{energy7})-(\ref{entropy7})
are thus simultaneous equations for $T$, $\dot{R}$.


The results of the numerical calculations can be
parametrized in the following form for the hadronic phase:
\begin{eqnarray}\label{eq:TandV}
V(\tau) &=&\pi\left[R + v\left(\tau-\tau_{c}\right)+a/2\left(\tau-\tau_{c}\right)^{2}\right]^{2}\tau c,
\nonumber \\
T(\tau) &=& T_{c}-\left(T_{h}-T_{f}\right)\left(\frac{\tau-\tau_{h}}{\tau_{f}-\tau_{h}}\right)^{\alpha},
\label{vol-param}
\end{eqnarray}
with $T_{h}$ and $T_{f}$ being the hadronization and kinetic
freeze-out temperature. We take $T_{h}$ = 156 ${\rm MeV}$ to be the same
as the cross over temperature $T_{c}=T_{h}$. Eq. \eqref{eq:TandV}
can be thought to describe the system of the hadronic phase of the nuclear matter expanding with transverse velocity $v$ and transverse acceleration
$a$ starting from its transverse size $R$ at the chemical
freeze-out time $\tau_{c}$. The values of $T_{f}$ according to
centrality are taken from Ref. \cite{Acharya:2019yoi}. The values
used in Eq. \eqref{eq:TandV} are summarized in Table
\ref{table:centrality}.
The schematic hydrodynamics in Eqs.~(\ref{energy7}) - (\ref{gamma}) assumes the expansion of a uniform matter whose temperature or particle density is the same anywhere in the local rest frame.
In the Lab. frame, however, the particle density increases with the radial distance due to Lorentz contraction.
We note that Eq.~(\ref{vol-param}) is not the parametrization for the volume in the Lab frame but for the volume with the Lorentz contraction taken into account, that is, $A\tau \langle \gamma_r\rangle$ in Eq.~(\ref{entropy7}).

\begin{table}
\caption{Values for the volume and temperature profiles in the phenomenological model Eq.\eqref{eq:TandV}.}
\centering
 \begin{tabular}{||c c c c c c c c||}
 \hline
Centrality & $T_{f}$ & $t_{c}$& $t_{f}$ & R  & v  & a & $\alpha$
\\
(\%)  &(MeV)&(fm/c) & (fm/c) &(fm) & (c) & ($c^{2}/fm$) &
  \\ [0.5ex]
   \hline\hline
$0-5\%$ &90& 8.7 & 28.1 & 14.9 & 0.93 & 0.04 & 0.835
 \\ \hline
 $40-50\%$ &108 & 4.9 & 13 & 7.8 & 0.78 & 0.052 & 0.9
 \\ \hline
  $70-80\%$ & 147  & 2.2 & 2.9 & 4.43 & 0.481 & 0.161 & 0.847
 \\ \hline
\end{tabular}
\label{table:centrality}
\end{table}

\section{The abundance of $K_{1}$ mesons in heavy-ion collisions}

\begin{figure}[!t]
\begin{center}
\includegraphics[width=0.45\textwidth]{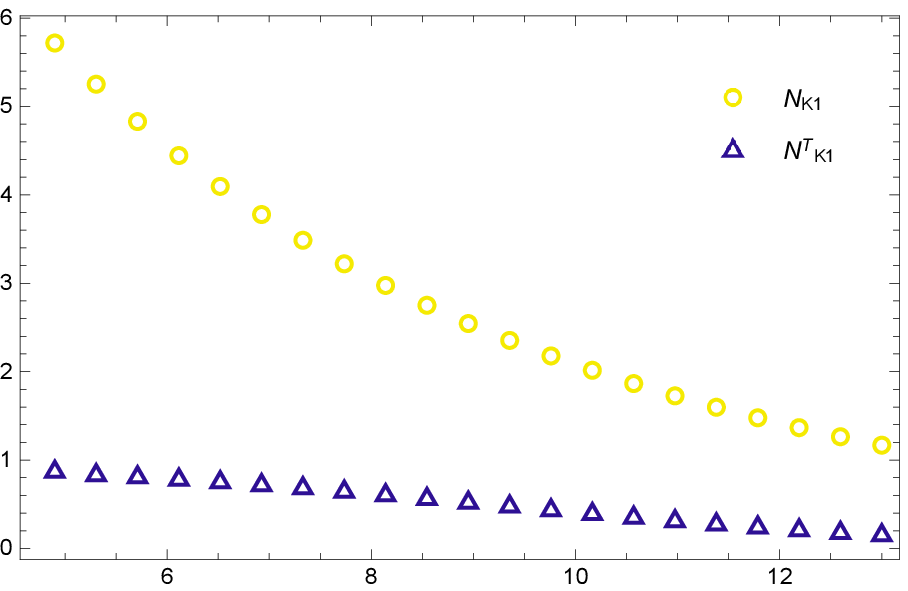}
\includegraphics[width=0.45\textwidth]{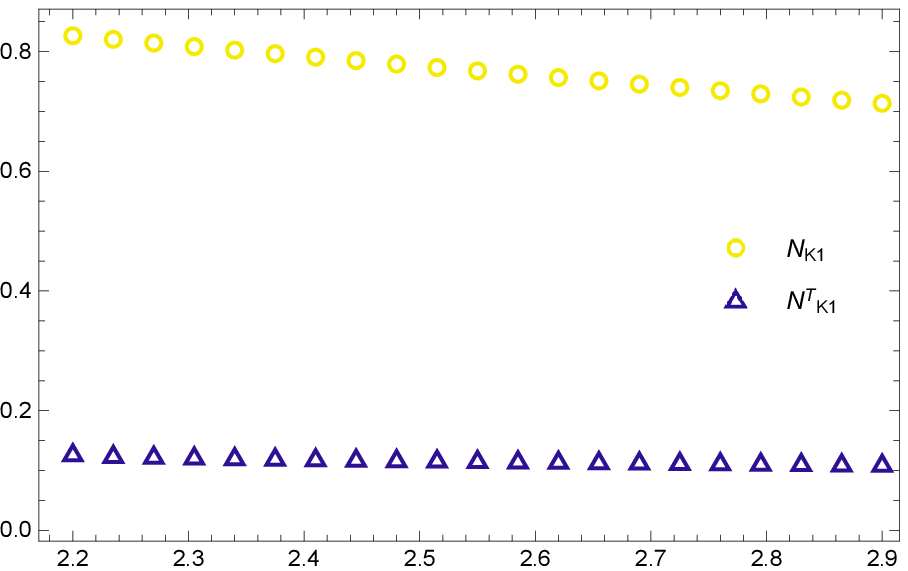}
\end{center}
 \caption{The time evolution of the $K_{1}$ meson abundance under hadronic interaction when the initial number is equal to $N_{K*}$(circle) or $N_{K_1}^T$(triangle) at $T_c$ for 40-50\%(upper graph) and 70-80\%(lower graph) centrality ranges. }
  \label{fig:NK1}
\end{figure}

\begin{figure}[!t]
\begin{center}
\includegraphics[width=0.45\textwidth]{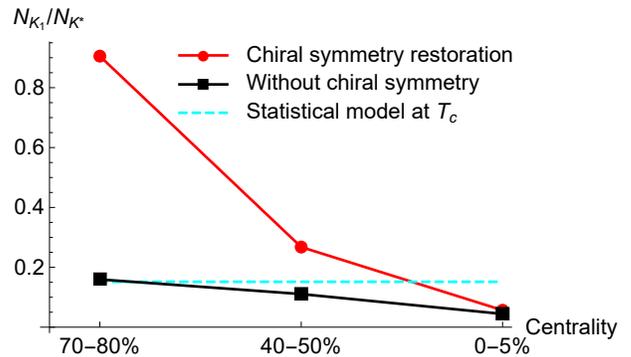}
\end{center}
 \caption{
The yield ratio of $K_1$ over $K^*$ with and without chiral symmetry
restoration for 70-80\%, 40-50\%, and 0-5\% centrality ranges. $K_1$ feed down to $K^*$ should be added when comparing to experiment.
 }
  \label{fig:centrality}
\end{figure}

Table \ref{table:KandKS} shows the abundance of the strange mesons
at the chemical  freeze-out point for different centrality ranges.
The yield ratio between the $K_1$ and $K^*$ mesons at the kinetic
freeze-out point is also given. $N_{K_1}$ is the solution of the
rate equation Eq.~(\ref{eq:rate}) with the initial number equal to
$N_{K^*}=N_{K*}^T$ at $T_c=156$ MeV in the presence of chiral
symmetry restoration. On the other hand, $N_{K_1}^T$ is the
expected yield without chiral symmetry restoration  when  the
initial number is given by its thermal equilibrium  number at
$T_c$. Hence, $N_{K_1}^T$ at the kinetic freeze-out point is the
number expected in the statistical model when hadronic
dissociation is taken into account.   For a clear signature of
chiral symmetry restoration at $T_c$, we want $N_{K_1}/N_{K^*}$ at
the kinetic freeze-out temperature to be sufficiently larger than
$N^T_{K_1}/N_{K^*}$ at $T_c$, which is the standard statistical
model prediction for the ratio.
  One notes that this is true for 40-50 \% and
70-80 \% centrality ranges.  In fact, as can be seen in the last
column of Table \ref{table:KandKS}, for  70-80 \%, one finds that
$N_{K_1}/N_{K^*}$ at the kinetic freeze-out point is 0.905, which
is much larger than $N^T_{K_1}/N_{K^*}$ at any time for all
centralities: the time dependencies of $N^T_{K_1}$ and $N_{K^*}$
are given in Fig. \ref{fig:NK1}. Fig. \ref{fig:centrality}
summarizes the yield ratios between $K_1$ and $K^*$ mesons for
different centralities with (red dots)  and  without (black
squares) chiral symmetry restoration at $T_c$. As can be seen in
the figure, the ratio increases sharply as the centrality
decreases (because of the higher kinetic freeze-out temperature
and thus the shorter lifetime of the hadronic phase at more
peripheral collisions \cite{Cho:2015exb, Motornenko:2019jha}) when
chiral symmetry restoration at $T_c$ is taken into account. It is
possible that at the chemical freeze-out point, the chiral order
parameter will partly acquire its vacuum value.  Then, the initial
ratio of $N_{K_1}/N_{K^*}$ at $T_c$ would be slightly less than 1.
Furthermore, the number of pions and rho mesons should be larger than estimated in Eq. \eqref{eq:density} due to the effective chemical potential during the hadronic phase\cite{Rapp:2000gy}, leading to  slightly larger dissociation cross sections than estimated in this work.
Nevertheless, the large increase in the  yield ratio towards
peripheral collisions would still be visible so that one can
identify chiral symmetry restoration. Hence, a systematic study of
the ratio $N_{K_1}/N_{K^*}$ for different centralities in a heavy
ion collision will unambiguously show the evidence of chiral
symmetry restoration at $T_c$.

\begin{table}
\caption{The kaon, $K^{*}$ and $K_1$  meson abundances under hadronic interaction at chemical and kinetic freeze out temperatures. $N_{K_1}^T$ is the number of $K_1$ meson assuming thermal equilibrium while $N_{K_1}$ is the abundance assuming chiral symmetry at $T_c$.  The last column shows our prediction when chiral symmetry is restored.  }
\centering
 \begin{tabular}{||c  c c c c c | c c||}
 \hline
Centrality & T(MeV) & $N_{K}$ & $N_{K^{*}}$& $N_{K_1}^{T}$ & $N_{K_1}$  & $\frac{N_{K_1}^T}{ N_{K^{*}}} $ & $\frac{N_{K_1}}{ N_{K^{*}}}$
  \\ [0.5ex]
   \hline\hline
$0-5\%$ & 156 & 80.5 & 37.0 & 5.62 & 37.0 & 0.152 & 1.00
 \\
  & 90 &   &  & 0.157 & 1.37 & 0.006 & 0.057
 \\ \hline
 $40-50\%$ & 156 &12.4 & 5.72 & 0.867 & 5.72 & 0.152 & 1.00
 \\
 & 108 &   &   &0.149 & 1.17  & 0.034 & 0.268
 \\ \hline
  $70-80\%$ & 156 &1.80 & 0.827 & 0.125 & 0.827 & 0.152 & 1.00
  \\
  & 147  &   &  &  0.108 & 0.714 & 0.137 & 0.905
 \\ \hline
\end{tabular}
\label{table:KandKS}
\end{table}

\begin{table}
\centering
\label{table:pion and rho}
\end{table}

\section{Summary }

We studied effects of chiral symmetry restoration at the chemical
freeze-out point in a relativistic heavy ion collision.  As the
mass differences between the chiral partners are order parameters
of the chiral symmetry restoration, the production of the chiral
pairs will be degenerate at the chemical freeze-out point, where
the chiral order parameter is found to be substantially reduced
from the vacuum value in lattice calculations. For the effects to
be observable, the vacuum width as well as the hadronic
dissociation of both particles should be small so that the signal
will not be smeared out during the hadronic phase. The vacuum
widths of both the $K_1$ and $K^*$ meson are smaller than 100 MeV
and thus potential candidates to be studied. We have performed a
systematic study on their hadronic absorption during the hadronic
phase as well as  on the centrality dependence
of these effects in a heavy ion collision. Our findings suggest
that while the anomalously large initial ratio between the $K_1$
and $K^*$ meson compared to that of the statistical model
prediction will most likely be smeared out in a central collision,
the signal will be visible in a peripheral heavy ion collision due
to the shorter life time of the hadronic phase and higher
freeze-out temperature.

The chiral partnership between the $K^*$ and $K_1$ exists between
the same charge states when the baryon chemical potential is
non-zero \cite{Song:2018plu}.  In ultra-relativistic heavy ion
collisions, the initial state will have almost zero baryon chemical
potential.  Hence, one can compare the production of any of the
charge states through the decay products given below or their
charge conjugation.  For the $K_1$ meson they are
\begin{align}
K_1^- \to \begin{cases}
\rho^0 K^-  \cr
\rho^- \bar{K}^0  \cr
\pi^0 K^{*-}  \cr
\pi^- \bar{K}^{* 0}
\end{cases}
, ~~~~
\bar{K}_1^0 \to \begin{cases}
\rho^+ K^-  \cr
\rho^0 \bar{K}^0 \cr
\pi^+ K^{*-}  \cr
\pi^0 \bar{K}^{* 0}  \end{cases}, \nonumber
\end{align}
and for $K^*$
\begin{align}
K^{*-} \to \begin{cases}
\pi^0 K^{-}  \cr
\pi^- \bar{K}^{ 0} \end{cases}
, ~~~~
\bar{K}^{*0} \to \begin{cases}
\pi^+ K^{-}  \cr
\pi^0 \bar{K}^{ 0} \end{cases} .
\nonumber
\end{align}
A systematic study of the production of these states depending on the centrality will lead us
to identify chiral symmetry restoration in heavy ion collision.

\section*{Acknowledgments}

This work was supported by Samsung Science and Technology
Foundation under Project Number SSTF-BA1901-04. S. Lim acknowledge
support from the National Research Foundation of Korea (NRF) grant
funded by the Korea government (MSIT) under Contract No.
NRF-2008-00458. S. Cho was supported by the National Research
Foundation of Korea (NRF) grant funded by the Korea government
(MSIT) (No. 2019R1A2C1087107).

\pagebreak

\end{document}